\documentstyle[11pt,aaspp4]{article}

\begin{document}

\title{Evolutionary models for disk galaxies, a comparison with the
  observations up to intermediate redshifts ($z\lesssim 1)$}

\author{Claudio Firmani\altaffilmark{1}} \affil{Centro de
  Instrumentos-UNAM, A.P. 70-186, 04510 M\'exico D. F., M\'exico; and}
\affil{Instituto de Astronom\'\i a-UNAM, A.P. 70-264, 04510 M\'exico
  D. F., M\'exico}

\and \author{Vladimir Avila-Reese\altaffilmark{2}} \affil{Instituto de
  Astronom\'\i a-UNAM, A.P. 70-264, 04510 M\'exico D. F., M\'exico}

\altaffiltext{1}{Email address: firmani@astroscu.unam.mx}
\altaffiltext{2}{Email address: avila@astroscu.unam.mx}

\begin{abstract}
  We present a scenario for the formation and evolution of disk
  galaxies within the framework of an inflationary cold dark matter
  universe, and we compare the results with observations ranking from
  the present-day up to $z\sim 1$. The main idea in this scenario is
  that galactic disks are built-up inside-out by gas infall with an
  accretion rate driven by the cosmological mass aggregation history
  (MAH). In Avila-Reese et al. (1997) the methods to generate the MAHs
  of spherical density fluctuations from a Gaussian random field, and
  to calculate the gravitational collapse and virialization of these
  fluctuations, were presented. Assuming detailed angular momentum
  conservation during the gas (5\% of the total mass) contraction, a
  disk in centrifugal equilibrium is built-up within the forming dark
  matter halo. The primordial angular momentum is estimated
through the Zel'dovich approximation and normalized to the spin parameter $%
\lambda $ given by analytical and numerical studies. The disk galactic
evolution is followed through a physically self-consistent approach
which considers (1) the gravitational interactions among the dark
halo, the stellar and gas disks, and a bulge; (2) the turbulence and
energy balance of the interstellar medium; (3) the star formation
process due to gas disk gravitational instabilities; and (4) the
secular formation of a bulge due to the gravitational instabilities of
the stellar disk.

We find that the main disk galaxy properties and their correlations
are basically established by the combination of three fundamental
physical factors: the mass, the MAH, and the spin parameter $\lambda
$. Models calculated for a statistically significant range of values
for these factors predict nearly exponential disk surface brightness
profiles with realistic central surface brightnesses $\mu _{B_0},$ and
scale lengths (including low surface brightness galaxies), nearly flat
rotation curves, and negative gradients in the B-V color index radial
distribution. The main trends across the Hubble sequence of the global
intensive properties such as B-V, $\mu _{B_0},$ the gas fraction
$f_g$, and the bulge-to-disk ratio b/d, are reproduced. For a given
mass (luminosity) B-V correlates with the maximum circular velocity,
and this correlation is in agreement with the scatter of the
Tully-Fisher relation.  We interpret the observed color-magnitude, and
``color '' Tully-Fisher relations as a result of the empirical
dependence of extinction on luminosity (mass). The model properties
tend to form a biparametrical sequence, where B-V and $\mu _{B_0}$
could be the two parameters. The star formation history depends on the
MAH and on the $\lambda$ parameter. A maximum in the star formation
rate for most of the models is attained at $z\sim 1.5-2.5$, where this
rate is approximately 2.5-4.0 times larger than the present one. The
scale radii and the bulge-to-total ratio decrease with $z$, while $\mu
_{B_0}$ increase. The B-band TF relation remains almost the same at
different redshifts. Our scenario of disk galaxy formation and
evolution reveals that the cosmological initial conditions are able to
determine the main properties of disk galaxies across the Hubble
sequence and predict evolutionary features for the present-day
dominant galaxy population that are in agreement with very
recent deep field observational studies
\end{abstract}


\section{Introduction}

The understanding of the formation and evolution of galaxies is one of
the clearest challenges of contemporary astrophysics and cosmology.
Since galaxies are both cosmological and astronomical objects, two
general approaches can be used in order to study their formation and
evolution (e.g., Renzini 1994): ({\it i}) the deductive approach,
through which, starting from some initial conditions given by a theory
of cosmic structure formation, one tries to follow the evolutionary
processes until the reconstruction of the observable properties of the
galaxies and ({\it ii}) the inductive approach, in which, starting
from the present-day properties of galaxies, and through galactic
evolutionary models, one tries to reconstruct the initial conditions
of galaxy formation; the increasing observational data on galaxies at
intermediate and high redshifts will enrich this approach with crucial
constraints.

Most of current theories about cosmic structure formation are based on
the gravitational paradigm and on the inflationary cold dark matter
(CDM) cosmological models. Since these models predict more power for
the small density fluctuation scales than for the larger ones, cosmic
structures build up hierarchically, through a continuous aggregation
of mass. From the point of view of the galaxy cosmogony, a crucial
question is whether this aggregation occurs through violent mergers of
collapsed substructures and/or through a gentle process of mass
aggregation. This question depends on the statistical distribution of
the density fluctuation field and on its power spectrum.
Nevertheless, even if the dark matter structures assemble through
chaotic and violent mergers of subunits, the baryon gas, because of
the reheating due to the shocks implied in the collapse, virialization
and star formation (SF) feedback processes, will tend to aggregate
around the density peaks in a more (spatially) uniform fashion than
dark matter do it. Within the framework of the hierarchical clustering
theory, from the most general point of view, two could be the galaxy formation scenarios. In
one case, the main properties of galaxies, including those which
define their morphological types, are supposed to be basically the
result of a given sequence of mergers. This picture, that we shall
call the{\it \ merger scenario,} has been widely applied in
semianalytical models of galaxy formation where galaxies are
constructed from the cosmological initial conditions through
preconceived recipes (e.g., Lacey et al. 1993; Kauffmann, White, \&
Guiderdoni 1993; Cole et al. 1994; Kauffmann 1995, 1996, Baugh, Cole,
\& Frenk 1996). In the other case, the formation and evolution of
galaxies is related to a gentle and coherent process of mass
aggregation dictated by the forms of the density profiles of the
primordial fluctuations: galaxies continuously grow inside-out. We
shall call this picture, firstly developed by Gunn (1981,
1987), and by Ryden \& Gunn (1987), the e{\it xtended collapse scenario. }%
Since disk galaxies ($\sim 80\%$ of present-day normal galaxies) could
not have suffered major mergers due to the dynamical fragility of the
disks (T\'{o}th \& Ostriker 1992), the extended collapse scenario
results more appropriate to study their evolution.

According to the merger scenario, the bulges of spiral galaxies and
the elliptical galaxies arise from the mergers of galactic disks. A
natural prediction of this scenario is that spirals with small
bulge-to-disk ratios should have bulges older than those of spirals
with large bulge-to-disk ratios (e.g., Kauffmann 1996). As Wyse,
Gilmore, \& Franx (1997) have pointed out this does not appear
compatible with recent observational data (de Jong 1996a; Peletier \&
Balcells 1996; Courteau, de Jong, \& Broeils 1997). On the other hand,
if elliptical galaxies are the product of relatively recent mergers,
then a big dispersion is expected in their color-magnitude
relationship (but see Kauffmann 1996). Bower, Lucey, \& Ellis (1992)
showed that for the ellipticals in the Coma Cluster, this relationship
is extremely tight. Ellis et al. (1997) confirmed this result for
ellipticals in intermediate redshift clusters, up to $z\sim 0.6.$ The
merger scenario could also have serious difficulties from the
dynamical point of view: it is not conclusive if mergers of disks are
able to reproduce the high central phase-space densities of elliptical
galaxies (e.g. Hernquist 1993).

The inductive approach yields the possibility to
establish several constraints to the galaxy formation and evolution
processes. Galactic evolutionary models{\it } have shown that due to
the rapid disk gas consumption in SF, closed models are not able to
explain several properties of disk galaxies, as well as the wide range
of colors, gas fractions, etc. that galaxies present across the Hubble
sequence (e.g., Larson \& Tinsley 1978; Tinsley 1980; Larson, Tinsley,
\& Caldwell 1980; Kennicutt 1983; Gallagher, Hunter, \& Tutukov 1984;
Firmani \& Tutukov 1992, 1994). On the other hand it was shown that
the SF time scale in disk galaxies is not controlled by the initial
gas surface density (Kennicutt 1983; Kennicutt, Tamblyn, \& Congdon
1994). Hence, models where gas accretion is introduced are more
realistic. Gas accretion could also be necessary to maintain spiral
structure. In the case of open models, galaxy formation and galactic
evolution might be two related processes where the SF time scale is
driven by the gas accretion rate at which the disk is being built up.
Infall models of disk galaxy formation have been recently favored by
studies of our own Galaxy and nearby galaxies (see for references
Cay\'{o}n, Silk, \& Charlot 1996). Moreover these inside-out disk
formation models seem also to be in agreement with constrictions
provided by deep field observations (e.g., Bouwens, Cay\'{o}n, \& Silk
1997; Cayon et al.  1996; see also Section 4).

The gas infall rate in luminous galaxies may be controlled by the
global process of galaxy formation (cosmological accretion) and/or by
a self-regulated process of SF formation. This latter process proposed
by White \& Rees (1978) and White \& Frenk (1991) is commonly
applied in the merger scenario models. According to this mechanism, the
gas accretion rate is driven by the cooling of the hot gas corona
sustained by the supernova-injected energy. In the extreme situation
of instantaneous galaxy formation, supernova gas reheating, halo
self-regulated SF, and cooling flows (if the reheated gas was not
completely expelled out of the system) become the dominant processes
in regulating luminous galaxy evolution. However, the self-regulated halo SF
model suffers from some inconsistencies. As Nulsen \& Fabian
(1996) pointed out, supernova feedback over large scales occurs on
roughly the same time scales as the SF, not fast enough to tightly
regulate the SF rate. In the same way, if a disk forms, then the
self-regulating mechanism of SF will apply to the disk where other
dynamical conditions prevail (see Firmani, Hern\'{a}ndez, \& Gallagher
1996), and not to the halo system. Unless the disk-halo connection is
very effective, the SF in the disk will not be regulated by a balance
of energy between the supernova input and the halo gas cooling. On the
other hand, the X-ray gas corona predicted by the self-regulated SF
mechanism lacks observational support, at least for the most massive
galaxies for which the X-ray emission would have been above the
minimum detection limits of the Rosat and ASCA experiments.

The galactic infall models suggested by the inductive approach, are
consistent with the cosmological (deductive) extended collapse
scenario of galaxy formation and evolution. In Avila-Reese, Firmani,
\& Hern\'{a}ndez (1997, hereafter AFH), within the framework of a
standard CDM model, the MAHs corresponding to fluctuations of galactic
scales were generated from the statistical properties of a Gaussian
random field. After calculating the virialization of the fluctuations,
a range of realistic dark halo structures were obtained. Now, with the
aim to explore whether these cosmological initial conditions are able to
predict the evolutionary and observational disk galaxy properties and
their correlations, particularly those which go across the Hubble
sequence (HS), we shall construct a self-consistent and unified model
of disk galaxy formation and evolution in the cosmological context.
Within the framework of the extended collapse scenario and using the
galactic evolutionary models of Firmani et al. (1996), we shall study
the formation and evolution of disks in centrifugal equilibrium into
the evolving dark halos. In section 2, the methods we use are
described. The model results at $z=0,$ the main predictions of the
models, and the comparisons with observations as regards the local (\S
3.1) and global galactic properties and their main correlations (\S
3.2) are presented in section 3. In section 4 we compare our
evolutionary models with observations at intermediate redshifts
($z\lesssim 1)$. Finally, the concluding remarks are given in section
5.

\section{The Method}

The first part of our approach is to calculate the gravitational
collapse and virialization of isolated dark galactic halos starting
from a primordial density fluctuation field. This was done in AFH and
the reader is referred there for details (see also Avila-Reese 1998).
We have assumed a Gaussian (random-phase) fluctuation field, for which
all the statistical properties depend only on the power spectrum of
fluctuations provided by the assumed cosmological model. The
conditional probability for Gaussian random fields is used to
calculate the mass distribution of halos at time $t_{i+1}$ that go on
to form halos of mass $M_i$ at a later time $t_i$. Fixing $M_0$ and
its cumulative density contrast at the present epoch $t_0,$ this
distribution is recurrently applied through a Monte Carlo method to
construct the MAHs of main progenitors. The MAHs were constrained to
simulate isolated objects. Assuming spherical symmetry, a generalized
secondary infall method was used to calculate the sequential collapse
of concentric shells given by the MAH, and their ``relaxation'' into a
stationary virialized structure. The structures obtained through our
method for the standard CDM (SCDM) model are in agreement with the
results of high-resolution cosmological N-body simulations. It was
also shown that, for a given mass the statistical dispersion in the
MAHs actually produces a sequence of different DM virialized
configurations.

\subsection{Disk build-up}

Gas is able to cool and further collapse within the DM halo. The
collapse can be halted only by centrifugal forces and/or when gas is
transformed into stars. While the second case could be related to the
origin of large bulges and elliptical galaxies, the former is commonly
invoked to explain the formation of disk galaxies. The build-up of
disks within the evolving DM halos is carried out as follows: (1) we
consider that, at the beginning, baryon matter has mass and angular
momentum distributions similar to those of the DM; it is assumed that
the spherical shells are in solid body rotation; (2) once the current
mass shell has attained its maximum expansion radius, a fraction $f$
of its mass (gas) is transferred in a dynamical time to a disk in
centrifugal equilibrium with a surface mass distribution calculated
under the assumption of detailed angular momentum conservation; for
galaxies the gas cooling time is typically smaller than the dynamical
time (Silk 1977; Rees \& Ostriker 1977; White \& Rees 1978; Ryden \&
Gunn 1987); (3) the gravitational drag on the total system produced by
the central disk is calculated at each time with the adiabatic
invariant formalism (Flores et al. 1993; AFH).

The specific angular momentum $j$ acquired by each collapsing shell
during the linear regime is estimated using the Zel'dovich
approximation (e.g., White 1994):

\begin{equation}
  j\propto t_m^{1/3}M^{2/3}\propto (GMr_m)^{1/2}
\end{equation}
where $t_m$ is the time at the maximum expansion, $M$ the mass inside
the shell, and $r_m$ the shell radius at the maximum expansion. The
constant in (1) is given by the amplitude of the tidal torque induced
by neighboring objects; obviously our approach is not able to predict
it. We use the predictions of analytical studies and the outcomes of
cosmological N-body simulations which are commonly expressed through
the spin parameter $\lambda \equiv \frac{J\left| E\right|
  ^{1/2}}{GM^{5/2}}$, where $J$ and $E$ are the angular momentum and
total energy of the virialized system (e.g., Catelan \& Theuns 1996
and references therein). These studies find that the distribution of
$\lambda ^{\prime }s$ is well approximated by a log normal
distribution with an average value of $\sim $0.05 and a width in the
log less than one. It was found that the dependence of this
distribution on the mass and power spectrum is weak. Therefore, we
will not consider it further.  We fix the constant needed in relation
(1) in such a way that the $\lambda $ of the virializing DM halo at
every time is equal to the value we chose from the distribution
mentioned above.

\subsection{Disk galactic evolution}

The disks which are continuously forming into the dark halos transform
their gas into stars. In Firmani et al. (1996), a self-consistent
approach to study the SF in its stationary regime, and the physics of
the gas and star disks was presented. In this approach gas disk height
and SF are regulated by an energy balance between supernova input and
turbulent energy gas dissipation. Here we add the kinetic energy input
due to gas accretion. Star formation runs when the Toomre
gravitational instability parameter for the disk gas ($Q_g\equiv
\frac{v_g\kappa }{\pi G\Sigma _g},$ $\kappa $ is the epicyclic
frequency, $v_g$ the gas velocity dispersion, and $\Sigma _g$ the gas
surface density) is below a given threshold, {\it i.e.} the SF is
controlled by a feedback mechanism such that, when the disk is
overheated by the SF activity, SF is inhibited and the disk
immediately dissipates the excess energy to lower $v_g$ back to the
value determined from the Toomre criterion. In his original study
Toomre (1964) estimated for an infinitely thin disk a value of 1 for
the threshold. Numerical simulations (Sellwood \& Carlberg 1984;
Carlberg 1985; Gunn 1987), and observational estimates (e.g., Skillman
1987; Kennicutt 1989) suggest thresholds of the order of 2; this
result is attributed to collective phenomena such as the swing and
``waser'' amplifications which are difficult to account for in the
analytical studies.  In our models the value of this threshold
controls the thicknesses of the gas and stellar disks; when a value of
2 is used we obtain for a model of the Galaxy gas and stellar
thicknesses compatible with observations. The gas loss from stars is
also included. The gravitational dynamics of the evolving disk, the DM
halo, and the bulge (see below) are treated in detail. The local disk
galactic models presented in Firmani \& Tutukov (1994) included an
integration over the evolutionary tracks of all stars formed (a
Salpeter initial mass function with a minimal star mass of
0.1M$_{\odot }$, and solar metallicities were used), allowing the
calculation of luminosities in different colors. Using the results
from these models the surface B-brightness and B-V color indexes may
be found at every radius and throughout the evolution of our galactic
disk models (see Firmani et al.  1996). We have compared the color
indexes obtained by Firmani \& Tutukov (1994) with those predicted by
the population synthesis of Fritze-v.Alvensleben \& Gerhard (1994) for
three SF histories that they identify as corresponding to Sa, Sb, and
Sd galaxy types. The agreement is satisfactory. We find that the
approximation to the population synthesis we use provides a B-V which
is $\sim 0.1$ mag in the red, and $\sim 0.05$ mag in the blue less
than the respective values given by more sophisticated models (e.g.,
Bruzual \& Charlot 1993; Charlot, Worthey, \& Bressan 1996).

\subsection{Bulge formation}

Recent observational and theoretical studies are changing some common
preconceptions about galactic bulges; for a recent review see Wyse et
al.  (1997). These studies tend to show that several bulge formation
mechanisms could be working in galaxies. Bulges in galaxies with low
and intermediate bulge-to-disk (b/d) ratios may be formed through
secular dynamical evolution of the disks, whereas bulges in galaxies
with large b/d ratios could have been formed separately from the
disks, through an early dissipative collapse and/or from mergers.

We have introduced a recipe to estimate the bulge mass. The stars of
the
central ``cold'' disk region where the stellar Toomre instability parameter $%
Q_s\equiv \frac{v_s\kappa }{3.36G\Sigma _s}$ is less than 1, are
transferred to a spherical component in such a way that $Q_s$ remains
equal to 1. The physical sense of this recipe is in agreement with the
secular scenario of bulge formation where gravitational instabilities
in the stellar disks produces bars which dissolve forming a ``hot''
component, the bulge (see Norman, Sellwood, \& Hasan 1996, and
references therein). Indeed, the similarity found in colors (Peletier
\& Balcells 1996), and scalelengths (de Jong 1996a; Courteau et al.
1997) between disk and bulge might mean that these parameters are
closely associated. It is worth emphasizing that our recipe of bulge
formation is a very crude approximation to a complex phenomena about
which, in fact, not much is known, both from the theoretical and
observational point of views.

\section{The models at z=0}

With the methods and schemes described in the previous section it is
possible, from the cosmological initial conditions, to calculate the
properties of disk galaxies at every time, particularly at the present-day, $%
z=0$. Here our purpose is to obtain {\it (i) }the main structural and
luminosity characteristics of galaxies corresponding to a local disk
galaxy population, and {\it (ii)} its main correlations between global
properties, particularly those which go across the HS. Once the
cosmological model and the present-day mass of the object under study
is given, the MAHs and their statistical distribution can be
calculated through Monte Carlo simulations.  To form disks in
centrifugal equilibrium it is necessary to fix the spin parameter
$\lambda $, and the fraction of total mass $f$ which is incorporated
in the disk. As in AFH, here we shall assume that galaxies incorporate
to the disk all the available baryons in the form of gas, and with a
settled mass fraction equal to the primordial Big Bang
nucleosynthesis predictions, $f_B\approx 0.05\Omega _oh_{0.5}^2,$ where $%
h_{0.5}$ is the Hubble parameter in units of 50 kms$^{-1}$Mpc$^{-1}.$
This assumption may not be realistic particularly for low mass
(velocity) systems which lose much of gas after the first bursts of
SF, giving rise to the family of dwarf galaxies (Dekel \& Silk 1986).
To achieve the first purpose, it is enough to calculate models for a
significative range of masses, MAHs and $\lambda ^{\prime }s$, and
test if the obtained galaxy properties for these ranges are realistic.
As a first order approximation, we shall calculate models only for the
average cases of the distributions of the MAHs and $\lambda ^{\prime
  }s$, and for two statistically significant deviations from the
averages. Therefore, regarding the second purpose, only general trends
in the correlations among the global galaxy properties will be
obtained. As a matter of fact, the available observational galaxy
samples are not complete enough as to provide certain statistical
information on galaxy properties and their correlations. We consider
three representative masses (dark+baryon matter) for normal galaxies:
5$\times 10^{10}$M$_{\odot },$ 5$\times 10^{11}$M$_{\odot },$ and
5$\times 10^{12}$M$_{\odot }.$ For each mass three MAHs are selected,
the average, and two symmetrical deviations in such a way that roughly
80\% of all the MAHs are contained between them (see AFH). For each
mass, and for each MAH, we shall calculate models taking three values
for $\lambda :$ 0.035, 0.05, and 0.1. As in AFH, with the aim of
studying general behavior, only one representative cosmological model
will be used, namely the Gaussian SCDM model normalized to $\sigma
_8=0.57$. In AFH it was shown that the $\sigma _8=1$ SCDM model
produces the correct slope of the I- and H-band Tully-Fisher (TF)
relations, but the zero-point is too small. This disagreement with
observations disappears when $\sigma _8=0.57$ is used, a value
suggested by studies of masses and abundances of rich clusters of
galaxies (White, Efstathiou, \& Frenk 1993). It is worth remarking
that at galactic scales the power spectrum of the flat, {\it
  COBE-}normalized $\Omega _\Lambda =0.7,$ $h=0.65,$ CDM model is very
similar to that of the $\sigma _8=0.57$ SCDM model.

\subsection{Local properties of the disk galaxy models.}

In Figure 1(a) the surface B-brightness profiles of a 5$\times 10^{11}$M$%
_{\odot }$ model are plotted. The solid, dashed, and point-dashed
lines correspond to cases with $\lambda $ =0.05, $\lambda $ =0.035,
and $\lambda $ =0.100, respectively, where the thick lines are for the
average MAH, and the thin lines are for the early active MAH (only the
$\lambda $ =0.035 case is plotted) and the very extended MAH (only the
$\lambda $ =0.100 case is plotted). It is seen that the brightness
profiles are nearly exponential.  For other masses the situation is
the same. Although the scale radii and the central surface
brightnesses fall within the corresponding ranges allowed by
observations, we note that the radii and surface brightnesses of the
modeled disks tend to the upper and lower limits of these ranges,
respectively. If transfer of angular momentum from baryon matter to DM
is considered, then the disk scale lengths decrease giving smaller and
more concentrated disks.  The radii and surface brightnesses of models
with $\lambda =0.1$ compare well with those of low surface brightness
(LSB) galaxies. The models predict negative radial gradients in
colors, as is seen in Figure 1(b) where the radial B-V distributions
are shown for the same models of Figure 1(a).  Accurate
multiwavelength surface photometry studies of galaxies (e.g., de Jong
1995) confirm that galactic disks tend to be bluer at the periphery,
although these gradients are typically smaller than those predicted by
our models. Experiments where only a moderated angular momentum
redistribution in the infalling gas was introduced, show that the
color gradients decrease.

\begin{figure}
\plotone{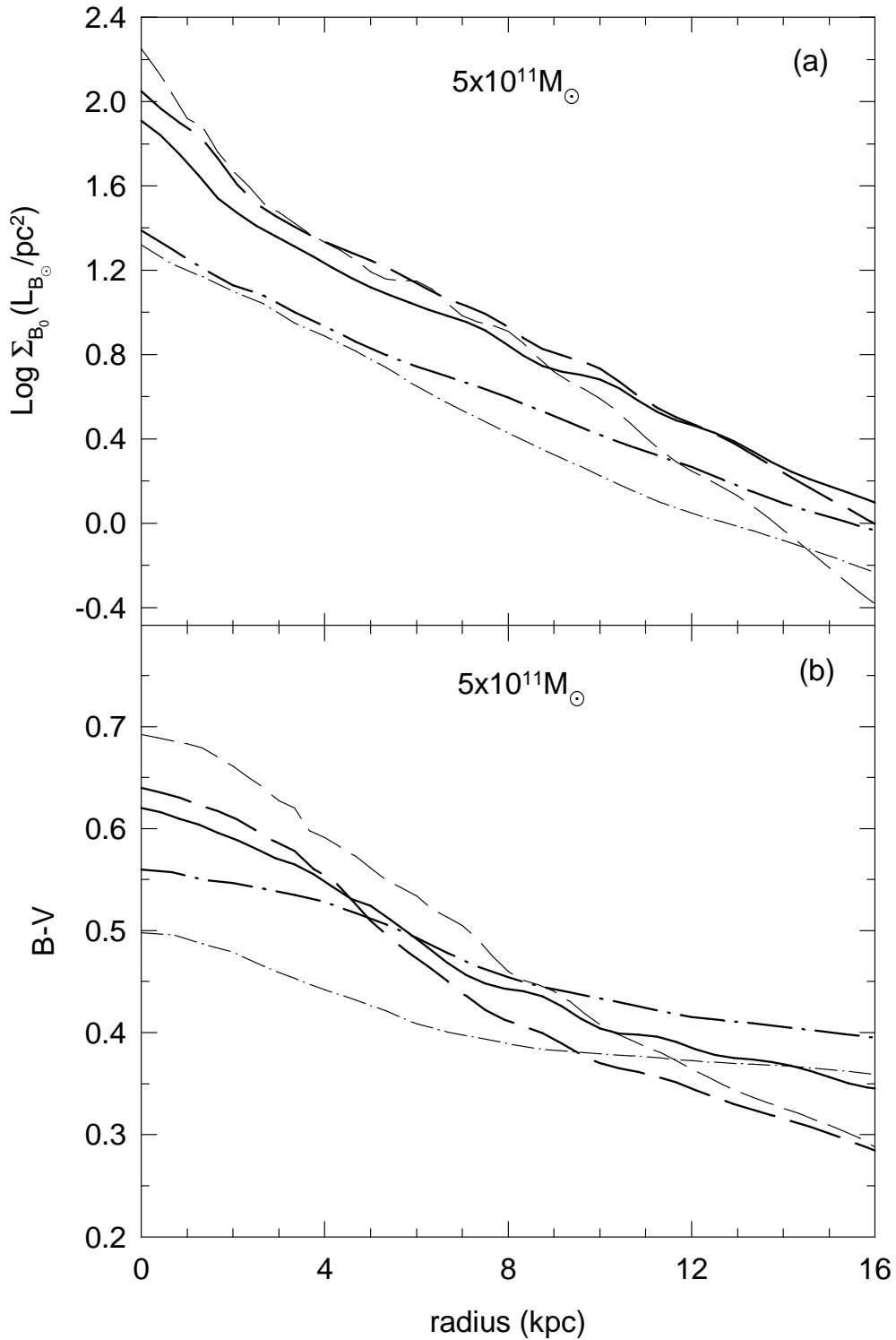}
\caption{The B-surface brightness (a) and the B-V color index
  (b) profiles of a 5$\times 10^{11}M_{\odot }$ galaxy. The average
  MAH cases for the spin parameters $\lambda =0.035$ (dashed line),
  $\lambda =0.050$ (solid line), and $\lambda =0.100$ (point-dashed
  line) are represented with the thick lines. For the early active MAH
  only the model with $\lambda =0.035 $ (thin dashed line) is plotted,
  while for the extended MAH, only the $\lambda =0.100$ case (thin
  point-dashed line) is shown. All the models were calculated for a
  $\sigma _8=0.57$ SCDM\ model. \label{fig01}}
\end{figure}

\placefigure{fig01}

In Figure 2 the rotation curves for the same models of Figure 1 are
depicted. The radii were scaled to the Holmberg radii. The shape of
the rotation curve depends mainly on the spin parameter $\lambda $ and
correlates strongly with the central surface brightness. For $\lambda $%
=0.035 the disk is very concentrated and the rotation curves are
decreasing at the optical radii. The rotation curves corresponding to models with $\lambda =0.05$ are nearly flat at the Holmberg radius for all the MAHs. For $\lambda =0.1$ the rotation curves grow slowly and the disks are less
concentrated than in the cases of smaller $\lambda ^{\prime }s$. In
Figure 2, the rotation curves of the 5$\times 10^{10}$M$_{\odot }$ and
5$\times 10^{12}$M$_{\odot }$ models corresponding to the average MAHs
and $\lambda $ are also depicted. At the Holmberg radius the less
massive models have nearly flat rotation curves, while the more
massive galaxies present decreasing rotation curves. The synthetic
rotation curves derived from observations (normal galaxies) in Persic,
Salucci, \& Stel (1996) show a similar trend, although in this case
the less luminous galaxies present increasing rotation curves at the
optical ($\sim $Holmberg) radius.

\begin{figure}
\plotone{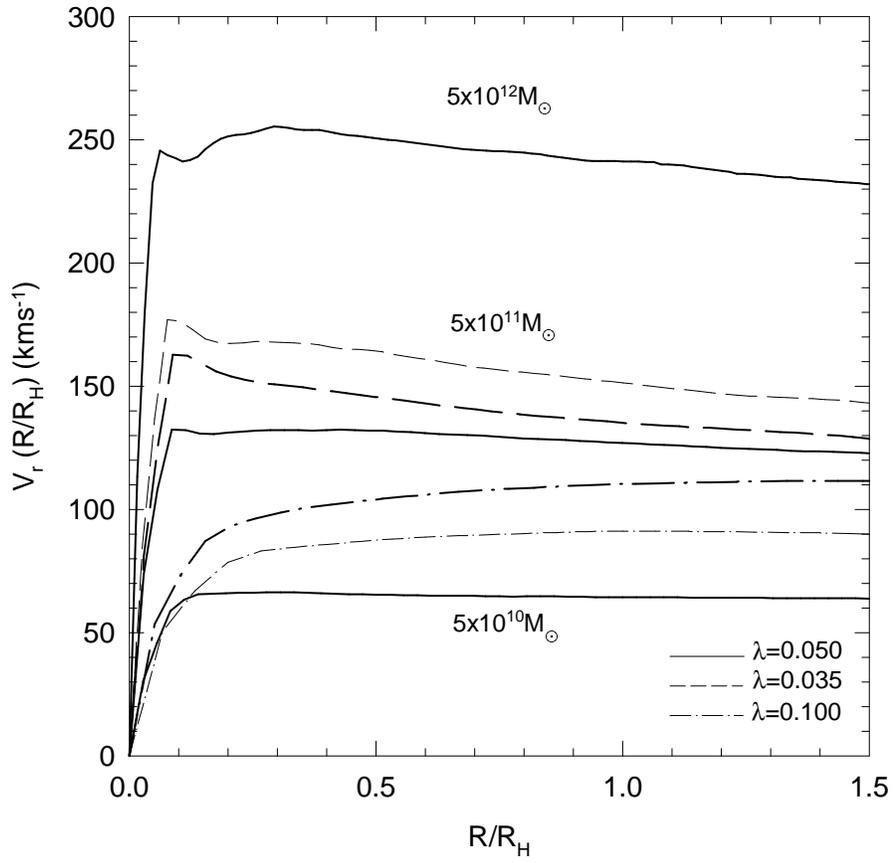}
\caption{Rotation curves for the same models of Figure 1 
(5$\times 10^{11}M_{\odot })$, and for a 5$\times 10^{10}M_{\odot }$ (bottom
  curve) and 5$\times 10^{12}M_{\odot }$ (top curve) galaxy
  corresponding only to the average MAH, $\lambda =0.050.$ Radii were
  scaled to the optical (Holmberg) radii of each model. \label{fig02}}
\end{figure}

\placefigure{fig02}

In AFH it was shown that the dark halo component typically dominates
down to near the center (see Fig. 4 of AFH) in the rotation curve
decomposition.  This possible disagreement of the models with
observations is related to the cuspy inner structure of dark matter
halos. Collective dissipative mechanisms and/or the cosmological
initial conditions could produce shallow cores in the dark halos
(AFH). Here, we shall artificially introduce constant density cores
according to the observed inner structure of dwarf and LSB\ galaxies.
Burkert (1995) found a density profile with two
parameters which fits very well the structure of dwarf galaxies: $\rho (r)=%
\frac{\rho _cr_c^3}{(r+r_c)(r^2+r_c^2)}$ . Actually, observations show
that the two parameters are correlated. Using the measured rotation
curves of the same five dwarf galaxies considered in AFH, we find that
$\rho _c=0.065\left( \frac{r_c}{kpc}\right) ^{-6/7}\frac{M_{\odot
    }}{pc^3}$ . The integration of this density profile until the
present-day halo virialization radius should be equal to the given
total halo mass $M_0$. The spherical top-hat collapse model and energy
conservation allow a rough estimate of
this radius as a function of the mass: $r_h\simeq 56\left( \frac{M_0}{%
  10^{10}M_{\odot }}\right) ^{1/3}h_{0.5}^{-2/3}$ kpc. Thus, using the
$\rho _c-r_c$ dependence, it is possible to estimate $r_c$ as a
function of $M_0$: $r_c\simeq 2.1\left( \frac{M_0}{10^{10}M_{\odot
    }}\right) ^{1/2}h_{0.5}^{-2/3}$ kpc. The Burkert density profile
also fits well the structure of LSB galaxies, although the dispersion
in the $\rho _c-r_c$ dependence is high. Given that the observational
information is very limited we shall use the presented mass-core
radius relationship only as representative of the average case. As is
seen in Figure 3 the rotation curve decomposition of a 5$\times
10^{11}M_{\odot }$ model with the average MAH, and $\lambda =0.035,$
which formed in a DM halo with a core, is similar to the
decompositions derived from the usual fitting techniques to
observations (e.g., Carignan \& Freeman 1985; van Albada et al. 1985;
Begeman 1987). Even galaxies with $\lambda =0.035$ now present a
nearly flat rotation curve. This allows us to shift the $\lambda
^{\prime }s$ to smaller values, within the uncertainty prediction
range, and in this way surface densities larger than in the coreless
case will be obtained. Furthermore, better agreement with the
synthetic rotation curves of Persic et al. 1996 is found. In
conclusion, the existence of cores in the DM halos produced in a SCDM
model, directly influences the dynamical and structural properties of
present-day disk galaxies, and in the correct direction.

\begin{figure}
\plotone{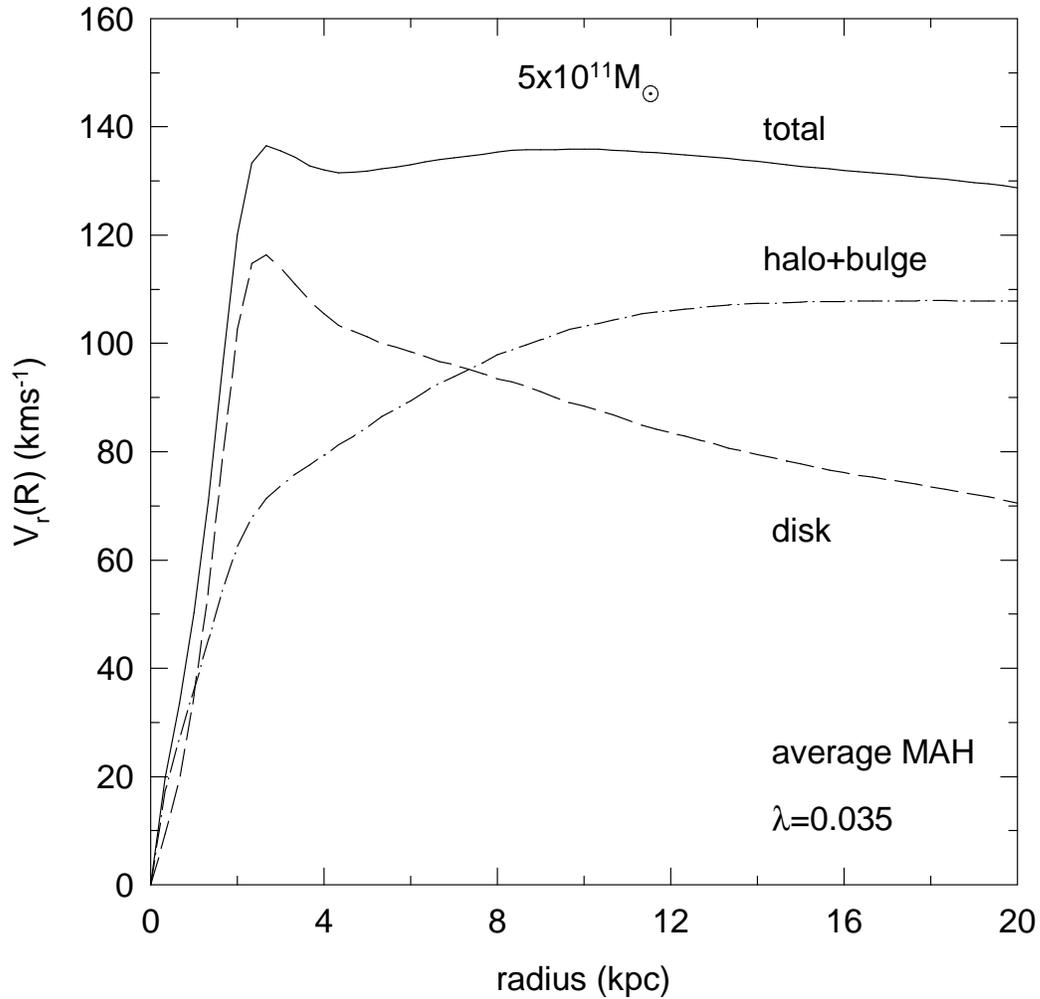}
\caption{Rotation curve decomposition of a 5$\times
  10^{11}M_{\odot }$ galaxy (average MAH and $\lambda =0.05$), in the
  DM halo of which an artificial near constant-density core was
  introduced. The size of the core was calculated in accordance with
  the observational data for the dwarf galaxies (see text). Compare
  this Figure with Figure 4 of AFH where the rotation curve
  decomposition of a galaxy without a core is presented. \label{fig03}}
\end{figure}

\placefigure{fig03}

\subsection{Global properties of disk galaxies and their correlations}

Among the global properties that our models predict for a disk galaxy
we shall consider the integral B-V color index (calculated within a
Holmberg radius), the B-band luminosity $L_B,$ the B-band exponential
disk scale length $h_d$, the B-band central surface brightness $\mu
_{B_o}$ or $\Sigma _{B_o}$ ($\mu _{B_o}$ is given in magnitudes per
arcsec$^2,$ and $\Sigma
_{B_o}$ in $L_{B_{\odot }}$ per pc$^{-2}$), the disk gas fraction $f_g$ ($%
\equiv \frac{M_{gas}}{M_{gas}+M_{stars}}$), the stellar bulge-to-disk
ratio b/d, and the maximum rotation velocity $V_{_{^{\max }}}.$ The
correlation matrix from a principal component analysis of these
properties for the 27 models calculated here is given in Table 1. It
is seen that for the intensive properties, the observational trends
across the HS are reproduced: the redder and more concentrated the
disk, the smaller is the gas fraction, and the larger is b/d.
Furthermore, the models seem to populate a planar region in the $\mu
_{B_o}-(B-V)-f_g$ and $\mu _{B_o}-(B-V)-b/d$ spaces, in rough
agreement with observations (McGaugh \& de Blok 1997, hereafter M-GB).
B-V and $\mu _{B_o}$ are almost independent one to another, so that we
can express $f_g$ and b/d as functions of these two parameters. In
figure 4, $f_g $ is plotted versus B-V and $\mu _{B_o}$. The gas
fraction is larger for smaller B-V (panel (a)) which means that $f_g$
is larger for the MAHs whose present-day gas infall rate is still
high. On the other hand, in panel (b) it is seen how a less
concentrated disk (larger $\lambda $) presents a higher gas fraction
than a disk with high $\Sigma _{B_o}$. On the basis of this result is
the influence of the disk gravitational compression on the capability
of gas to form stars (Firmani \& Tutukov 1992, 1994). The disk surface
density also strongly influences the b/d ratio because the stellar
surface density enters in the Toomre gravitational instability
criterion which is used in our models to calculate the formation of
bulges. The larger the central surface brightness (stellar density),
the larger is the b/d ratio (Figure 5 (b)). The mass and MAH introduce
a dispersion in this correlation because they influence the other
quantities which appear in Toomre criterion. In Figure 5(a) it is seen
how marginally b/d depends on
B-V. The observational data tend to confirm that b/d is more correlated to $%
\mu _{B_o}$ than to B-V (e.g., de Jong 1996a).

\placetable{tbl-1}

In Figures 4 and 5 are also plotted the observational data taken from
a compilation presented in M-GB where LSB galaxies are included (only
in Figure 4). The B-V color indexes were not corrected for the
internal (inclination) galaxy extinction. We have applied this
correction according to the formula given in the RC3 catalog (de
Vaucouleurs et al. 1991). The b/d ratios for the LSB galaxies
presented in M-GB were not estimated, so that in the panels where this
ratio is plotted the LSB galaxies are not considered. In general, the
models fall rather well within the observational ranges, and are in
agreement with the observable correlations (compare also Table 1 with
the correlation matrix presented in M-GB). It is surprising that the
b/d ratios predicted by the models using the simple gravitational
instability criterion are in agreement with those inferred from
observations (de Jong 1996a,b). Note that de Jong (1996b) used an
exponential profile in his two dimensional bulge-to-disk decomposition
procedure, instead of the de Vaucouleour's profile for the bulges,
arguing that such a profile fits the observations better. That is why
the b/d ratios obtained by him are smaller than those given by
previous b/d decompositions (e.g., Simien \& de Vaucouleurs 1986).

\begin{figure}
\plotone{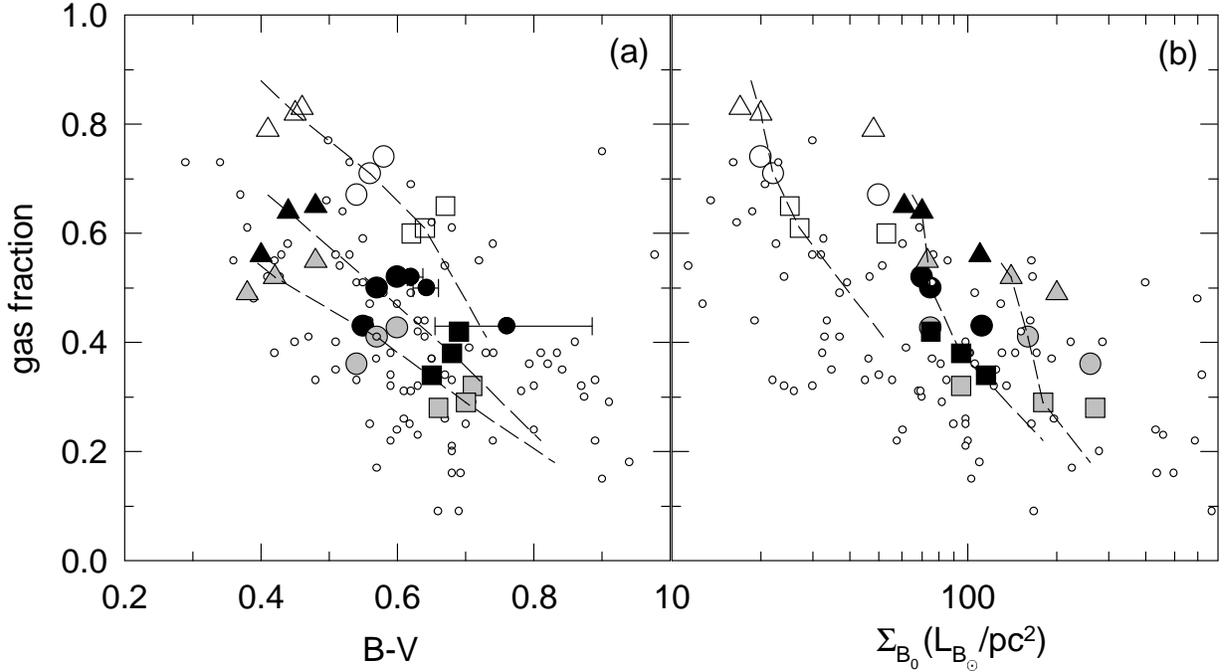}
\caption{The gas fraction $f_g$ vs. the integral B-V color 
index (a), and vs. the central B-surface brightness $\mu _{B_0}$ (b) for
  models and observations. The gray, black, and white filled symbols,
  correspond to models with $\lambda =0.035,$ $\lambda =0.050,$ and
  $\lambda =0.100$ respectively. Squares are for the early active MAH,
  circles for the average
MAH, and triangles for the extended MAH. Three masses (dark+baryon), 5$%
\times 10^{10}M_{\odot },$ 5$\times 10^{11}M_{\odot },$ and 5$\times
10^{12}M_{\odot }$ are considered (the larger the mass, the smaller is
the
gas fraction). The dashed lines connect the models of constant mass for 5$%
\times 10^{11}M_{\odot },$ and extend the statistical range of MAHs to
94\% (symbols consider only 80\% of the MAHs). The three small black
filled circles are the same models corresponding to the big black
filled circles but reddened according to the dust
absorption-luminosity dependence given in Wang \& Heckman (1996) (see
text). The error bars correspond to the range of values which fit
observational data. Small empty circles are the observational data
collected by McGaugh \& de Blok (1996) and corrected for inclination
extinction. LSB\ galaxies are included. \label{fig04}}
\end{figure}

\placefigure{fig04}

\begin{figure}
\plotone{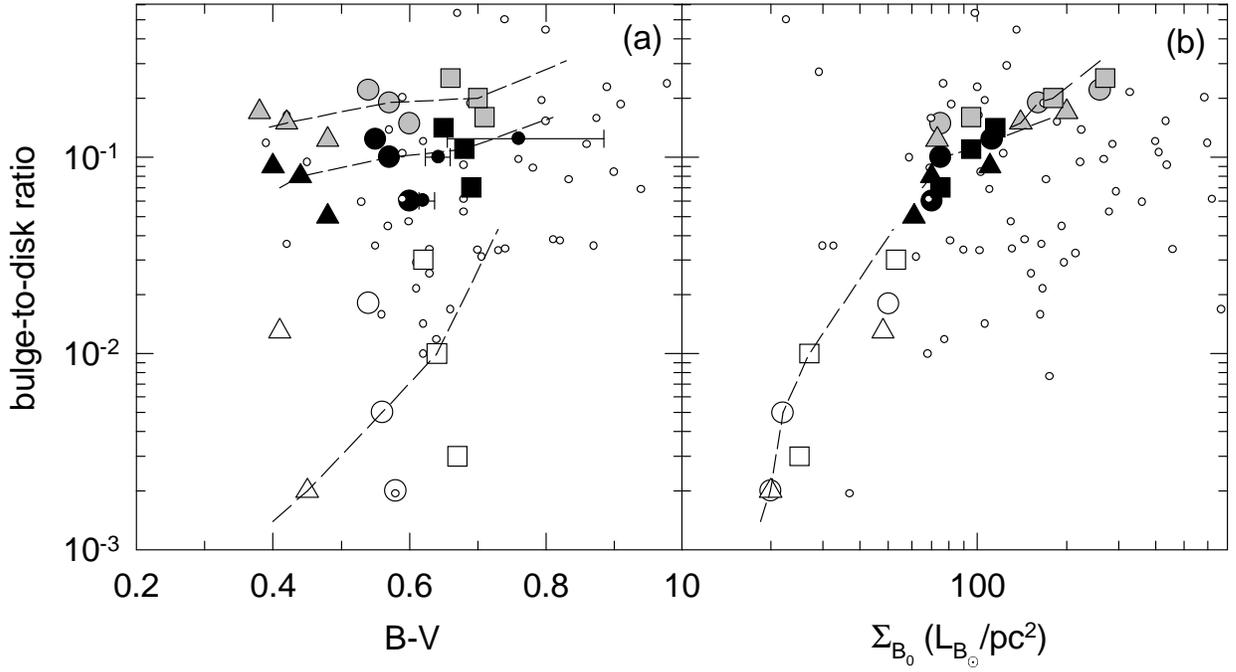}
\caption{The bulge-to disk ratio vs the integral B-V color index
  (a), and the central B-surface brightness $\mu _{B_0}$ (b) for
  models and observations. The same symbol and line codes of Figure 4
  are used. The bulge-to-disk ratios were taken from the K-band
  two-dimensional decompositions carried out by de Jong (1996b). LSB
  galaxies and a few normal galaxies shown in Figure 4 are absent in
  this Figure. \label{fig05}}
\end{figure}

\placefigure{fig05}

Perhaps the most serious inconsistency when comparing theory with
observation is that the considered models in Figures 4 and 5 do not
seem to be able to attain enough red color indexes. The statistical
range ($\sim 80\%)$ of the MAHs calculated here for the Gaussian
fluctuations leads to disks with B-V between $\sim 0.4$ and $\sim
0.7$. We have found that the color index becomes very sensitive to the
MAH when this corresponds to early high aggregation rates: for some
extreme cases the B-V color may be as red as $\sim 0.95$ mag. Thus,
some models can easily attain colors redder than 0.7; of course the
frequency of such models will be low. In Figures 4 and 5,
with dashed lines we show the range in the different properties of the 5$%
\times 10^{11}M_{\odot }$ models for the three $\lambda ^{\prime }s,$
when the statistical range in the MAHs is symmetrically extended to
94\% (symbols consider only 80\%). In the case of the color index it
is seen that the red extreme is very sensitive to the MAH. The Figures
show that roughly 3\% of models (5$\times 10^{11}M_{\odot })$ can be
redder than $0.8$ mag. The very incomplete observational sample we are
using here shows that $\sim 10\%$ of galaxies are redder than this
magnitude. A cross sample of 330 galaxies from the RC3 (de
Vaucouleours 1991), and the Tully (1988) catalogs (see Firmani \&
Tutukov 1994) would give $\sim 5\%.$ However, our aim in this
discussion is not to claim statistical predictions, for which there
would not be complete observational counterparts, but simply to point
out that our models can be as red as some observed galaxies are. On
the other hand, several questions not directly related to the scenario
presented here might be involved in the color index problem. (1) The
internal extinction in combination with the metallicity-luminosity
relation can introduce an important effect of reddening, particularly
for the most massive galaxies (see \S 3.2). Using the results
presented in Wang \& Heckman (1996), and the Galactic extinction curve
for $R_V=3.1$ (Cardelli, Clayton, \& Mathis 1989) we have reddened the
models corresponding to the average MAH and $\lambda =0.05$,
represented in Figures 4 and 5 with black filled circles. The error
bars account for the range of parameter values given in Wang \&
Heckman (see \S 3.2). (2) The influence of environment on the galaxy
evolution might help to produce models redder than 0.7 mag. In the
dense environments the gentle mass aggregation can be early truncated
and followed by a merging process between neighboring systems.
Experiments show that if gas accretion is truncated in the models at 6
Gyrs (4 Gyrs) then B-V roughly increases by 0.08 (0.15) mag. On the
other hand, the interactions in the dense environments can induce non
stationary SF which produces a fast gas consumption in stars. (3)
Although the statistical approach used to generate the MAHs has proved
to be a good approximation with respect to results of cosmological
N-body simulations for the average case (AFH), in the more extreme
situations of highly decreasing aggregation regimes the approach is
not necessarily realistic. In this approach, basically due to the
Gaussian statistics (where negative densities are possible), the mass
aggregation never stops. Hence, the possibility that the galaxy
neighborhood is matter exhausted is never taken into account, and as
was mentioned above B-V becomes very sensitive to the MAH in the cases
of highly decreasing aggregation regimes.

Although the observational sample presented in Figures 4 and 5 is
statistically incomplete, it seems that observations show larger
scatters in the correlations among the intensive properties than do
the models. While part of the scatter is produced by the observational
uncertainties and possible effects of the extinction, it is highly
probable that the intrinsic scatter is in any case larger than the one
predicted by the models, because the scenario proposed here does not
take into account several phenomena which are not dominant, but are
present in real galaxies. For example the SF prescription used in the
models is basically a stationary process, while in real galaxies SF
may appears in bursting modes. This fact introduces a stochastic
component in the photometric features, particularly in low mass
galaxies (Firmani \& Tutukov 1994).

With respect to the observations, the predicted central surface
brightnesses are smaller, while the gas fractions are slightly larger.
It is worth emphasizing that the scenario presented here assumes
detailed angular momentum conservation in the gas collapse, and all
the matter accreted by the disk is considered to be only in form of
gas. If some angular momentum transference from baryon matter to dark
matter is present during the gas collapse, and if some (small)
fraction of matter is incorporated to the galaxy in form of stellar
systems (mergers), then the modeled disks would be more concentrated
and less gaseous than those presented here.

The intensive properties and correlations of the models calculated
with an artificial core in the DM halo do not significantly differ
from models without a core. That is why we will not repeat Figures 4
and 5 for the models with a core. The most important change is related
to the shape of the rotation curves: in order to obtain flat rotation
curves for the average MAHs, $\lambda $ should be shifted to smaller
values than in the corresponding cases of coreless halos (see Figure
3), producing disks with slightly larger central surface brightnesses.

Concerning the extensive properties, the models predict close
relations between mass (luminosity) and maximum rotation velocity (the
Tully-Fisher (TF) relation), as well as between mass (luminosity) and
scale or Holmberg radius. In AFH it was shown that the disk mass vs.
maximum rotation velocity relation for the $\sigma _8=0.57$ SCDM\ 
model is in excellent agreement with the same relation estimated from
the observed H- and I-band TF relations using the appropriate
mass-to-luminosity ratios. It is interesting to note that the scatter
in the mass-velocity relation (and therefore probably in the H- or
I-band TF relations) is correlated with some intensive properties
which define the HS. For example, in Figure 6 it is shown how, for a
given mass, the maximum circular velocity increases with B-V. This
dependence can be easily understood from the point of view of the
extended collapse scenario: galaxies formed through gentle MAHs will
be less concentrated (smaller circular velocities) and with SF
histories more extended in time (bluer colors) than galaxies formed
through early active MAHs. The observational data confirm this
prediction of the models. In Figure 6, some galaxies from a cross of
the RC3 and Tully catalogs (see above) are also depicted. To estimate
the behavior of the rotation velocity with B-V for a given mass, the
data were divided in 3 bins according to the B luminosities presented
in the mentioned catalogs. The dashed lines are lineal regression to
each one these bins. The trend is roughly the same as in the case of
the models.

\begin{figure}
\plotone{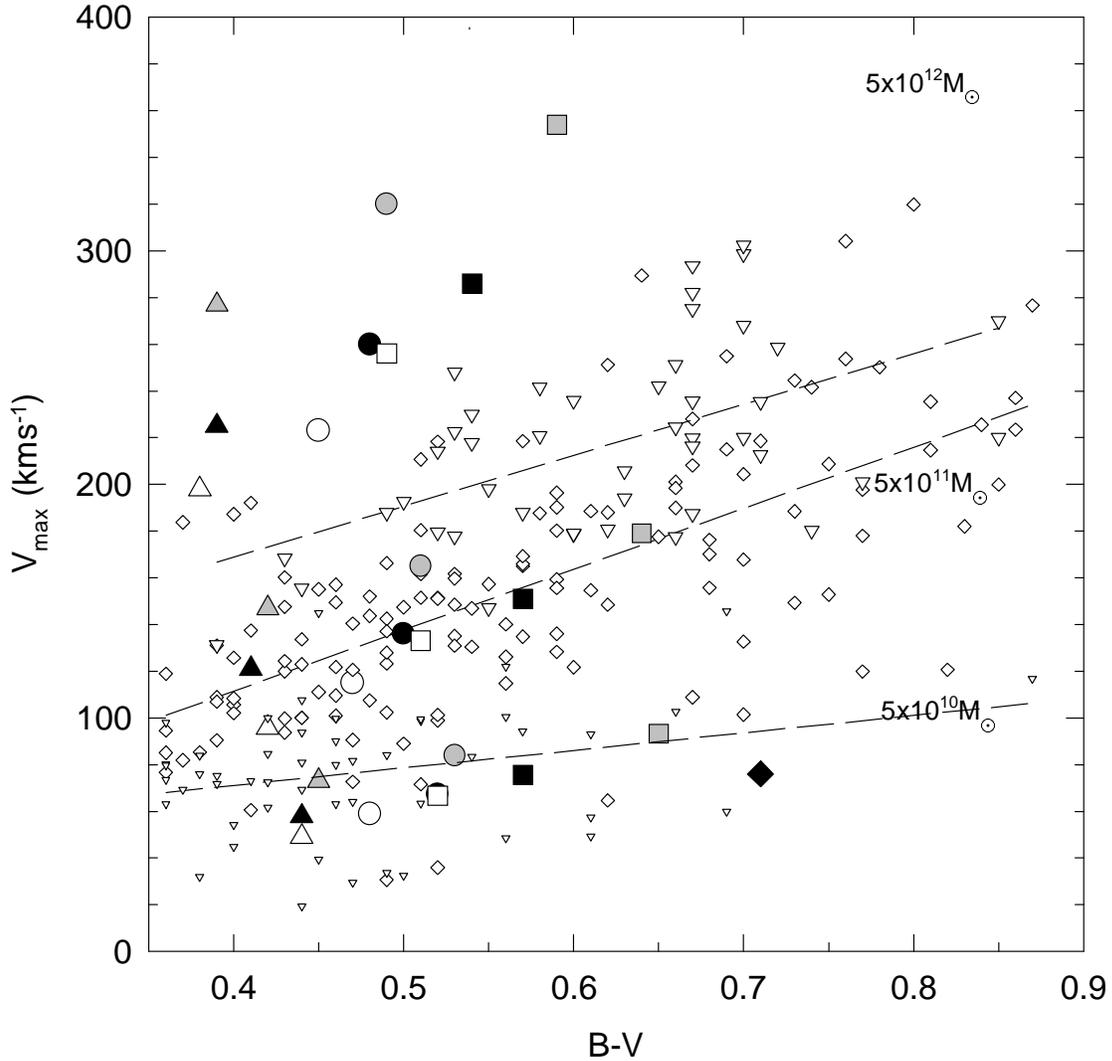}
\caption{The maximum rotation velocity vs. B-V for models and
  observations. The same symbol codes of Figure 4 are used. The
  observational data (small symbols) were taken from a cross of the
  RC3 and the Tully (1988) catalogs (see text). The small triangles,
  diamonds, and inverse triangles
correspond to galaxies with luminosities in B band within the 10$^8-3%
\times ${10}$^9L_{B_{\odot }},$ 3$\times ${10}$^9-3\times
10^{10}L_{B_{\odot }}$, and 3$\times 10^{10}-2\times
10^{11}L_{B_{\odot }}$ ranges, respectively. The dasehd lines are
linear regresions to the observational data corresponding to these
ranges. Note how the maximum velocity of models and observations for a
given mass (or range of luminosities) correlates with the B-V color. \label{fig06}}
\end{figure}

\placefigure{fig06}

The mass-radius relation predicted by the models where the average
values in the MAH and $\lambda $ were used is:

\begin{equation}
  M_s\propto R_H^{2.3}
\end{equation}
where $M_s$ is the disk stellar mass, and $R_H$ is the Holmberg
radius. The radius scales with the B-band luminosity $L_B$ as
$R_H^{2.4}.$ The Holmberg and scale radii do not correlate with any
intensive galaxy property (see Table 1), suggesting that the evolution
of disk galaxies and the HS are size independent (de Jong 1996a;
M-GB).

The models do not predict any correlation between B-magnitude (or
stellar
disk mass) and color, and the average B-band TF relation we obtain is $\frac{%
  L_B}{L_{B_{\odot }}}$=200$\left( \frac{V_{_{\max
      }}}{kms^{-1}}\right) ^{3.5}. $ Observational data point to a
correlation between magnitudes and colors (e.g., Vishnavatan 1981;
Wyse 1982; Tully, Mould, \& Aaranson 1982; Gavazzi 1993; Wang \&
Heckman 1996), and show slopes in the B-band TF relation smaller than
in the cases of the TF relations at longer wavelengths (for a review
see Strauss \& Willick 1995); these points are related to one another.
By dividing the B-band TF relation $L_B=A_BV_{\max }^{m_B}$ by, for
instance, the H-band TF relation $L_H=A_HV_{\max }^{m_H}$, one
obtains:

\begin{equation}
(B-H)=2.5(\frac{m_H}{m_B}-1)\log L_B+2.5\log \left( \frac{A_H}{A_B^{m_H/m_B}}%
\frac{L_{B_{\odot }}}{L_{H_{\odot }}}\right)
\end{equation}
For the values of $m_B$ and $m_H$ reported in the literature eq. (3)
gives (B-H) $\propto \alpha \log L_B$ with $\alpha \approx 0.4-1.2$ in
rough agreement with the reported magnitude-color relations. Within
the framework of the galactic evolutionary models presented here, the
{\it SF history} {\it is not able} to account for this dependence. It
is possible that the observed luminosity-metallicity relation and dust
extinction in galaxies are responsible for this dependence. In the
last few years the number of studies which point out to non-negligible
face-on extinction corrections have increased (see references in Wang
\& Heckman 1996, and Boselli \&\ Gavazzi 1994). Wang \& Heckman
(1996), based on studies of the far UV and FIR fluxes of a sample of
normal late type galaxies, have concluded that the dust opacity
increases with the luminosity of the young stellar population.
Through models of absorption and emission of radiation by dust in
uniform plane-parallel slabs they find that a power law optical depth
vs. UV luminosity relation explains this observational dependence.
This relation referred to the B band is:

\begin{equation}
  \tau _B=\tau _{B,*}\left( \frac{L_B}{L_{B,*}}\right) ^\beta
\end{equation}
where the best fits to observations are for $L_{B,*}=1.3\times
10^{10}L_{B_{\odot }},$ $\tau _{B,*}=0.8\pm 0.3,$ and $\beta =0.5\pm
0.2$.  According to the uniform slab model, the extinction in
magnitudes may be
expressed as $A_B=-2.5\log \left( \frac{1-\exp (-\tau _B)}{\tau _B}\right) $%
, and in the range $10^8-10^{11}L_{B_{\odot }},$ using (4) with the
central
values, is well approximated by $A_B\approx 0.38+0.42\log \left( \frac{L_B}{%
10^{10}L_{B_{\odot }}}\right) +0.14\left( \log \left( \frac{L_B}{%
10^{10}L_{B_{\odot }}}\right) \right) ^2$. Now, applying this
correction to the B-band luminosities given by our models, we are able
to predict the B-band TF relation influenced by the extinction. Figure
7 shows that this relation in the range $\sim 10^9-10^{11}L_{B_{\odot
    }}$ is well approximated by a line with slope $\sim $2.7, i.e. the
predicted TF relation now agrees with the observational estimates.
Note that the luminosity dependence of the extinction does not only
produce a change of slope, but also some nonlinearity, particularly at
the bright end of Figure 7. A similar result was previously reported
by Giovanelli et al. (1995), and this could be the reason for the
different slopes given for the B-band TF relation by different
authors. Recently, Kudrya et al. (1997) have presented the B-band TF
relation for a large sample of galaxies; from their Figure 6 it is
clearly seen how the slope of this relation tends to be steeper for
the less luminous galaxies. Since the intrinsic dust absorption and
the thickness of the dust layer with respect to a given stellar
population tend to decrease with increasing wavelength, the change of
slope in the TF relation will decrease as the passband tends to the
H-band. Thus, the measured dependence of dust extinction on luminosity
might explain part of the so called ``color'' TF relation rather well.
In Figure 7 are also depicted the corrected TF relations corresponding
to the two extreme cases of maximal and minimal optical depths, where
$\tau _{B,*}=1.1$ and $\beta =0.7,$ were used for the former, and
$\tau _{B,*}=0.5$ and $\beta =0.3$ for the latter. The more realistic
``sandwich'' model (Disney, Davies, \& Philipps 1989) was also
considered. While $\beta $ was not changed with respect to its
fiducial value, $\tau _{B,*}$ and the ratio of the height scale of
dust to young stars, $\zeta ,$ were fixed to the values suggested by
Bosselli \& Gavazzi
(1994) for the optically thin case in the H-band ($\tau _{B,*}=$1.33 and $%
\zeta =$0.74).

\begin{figure}
\plotone{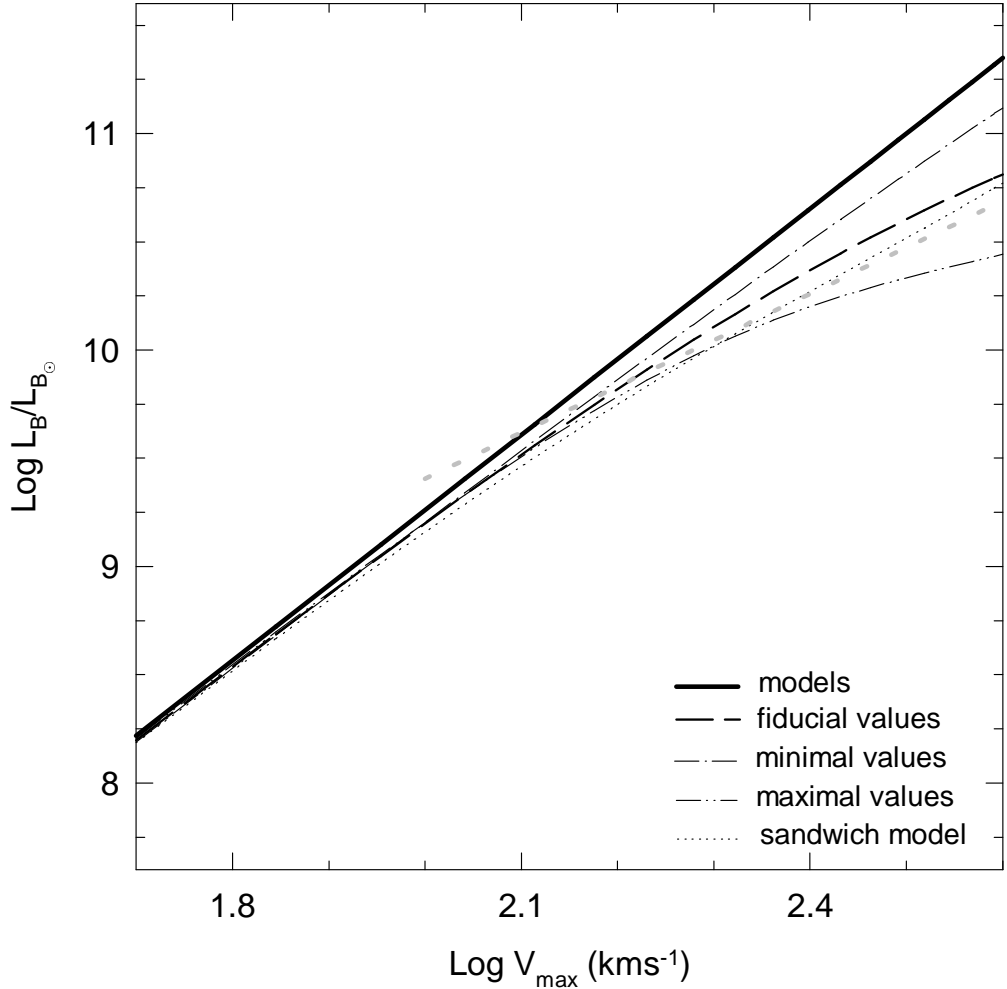}
\caption[fig07.eps]{The predicted B-band TF relation for the 
$\sigma _8=0.57$
  SCDM model (thick solid line). The slope of this relation is 3.5.
  The other lines show how the intrinsical TF relation transform if
  the B-luminosities are dimished by dust absorption according to the
  observational dependence of optical depth of dust on luminosity
  given in Wang \& Heckman (1996). While the dashed line corresponds
  to the fiducial optical depth, the point-dashed and two point-dashed
  lines are for two extreme cases of maximal and minimal optical
  depths (see text). The point line was obtained using a sandwich
  model with a ratio of height scale of dust to young stars of 0.74
  (see text for references). In the range $10^9-10^{11}L_{B_{\odot }}$
  the dashed line is well approximated by a line with slope $\sim
  2.7$. The dotted gray curve is the linear regresion to the empirical
  correlation between the absolute magnitude in B and the H21
  linewidth, $W_{50}^{corr}$, given in Kudrya et al. (1997). We have
  assumed $W_{50}^{corr}=2\times V_{\max }.$ We have truncated the
  regresion at $V_{\max }=100$ km/s because it does not provide a good
  approximation for lower velocities (see Figure 6 of Kudrya et al.
  1997). \label{fig07}}
\end{figure}

\placefigure{fig07}

Wang \& Heckman (1996) have pointed out that eq. (4) may be explained
by the observed increase in the metallicity with the luminosity for
which the dependence is considerable, $Z\propto L_B^\gamma ,$ with
$\gamma \approx 0.3-0.5$ (see for references Roberts \& Haynes 1994).
The metallicity also influences the spectrophotometric evolutionary
models. For example the models of Bressan, Chiosi, \& Fagotto (1994)
applied to single stellar populations with different initial
metallicities show that at $\sim 12$ Gyr
the differences in the V-K and B-V colors are $\sim $0.88 mag and $\sim $%
0.23 mag, respectively for a factor of 20 of variance in the
metallicity.  These models roughly agree with the observed
color-magnitude relation of elliptical galaxies, and in accordance
with Kodama \& Arimoto (1997), this relation is consequence of a
metallicity effect, instead of an age effect.  We conclude that given
the metallicity-mass relation observed in galaxies, extinction, and
the spectrophotometric evolution, may be the basis for the
color-magnitude and ``color'' TF relationships.

\subsection{The fundamental physical factors of disk galaxies and the Hubble
  Sequence}

The three main physical factors of the extended collapse scenario are the%
{\it \ total mass M, the MAH, and the angular momentum expressed
  through the spin parameter }$\lambda $; these parameters, and their
correlations are related to the initial cosmological conditions.
Analytical studies (Hoffman 1986, 1988;\ Heavens \& Peacock 1988)
suggest that $\lambda $ is almost independent on the fluctuation peak
height (related to the MAH) for CDM power spectra (but see Catelan \&
Theuns 1996). As a first approximation, here we consider that $\lambda
$ and the MAH are independent. We also assumed independence between
$\lambda $ and $M$ (see \S 2.3). Concerning the dependence of MAH on
$M$, for the CDM power spectra, the less massive galaxies show faster
early collapses than do more massive galaxies (AFH).  The correlation
coefficients from a principal analysis of these 3 factors with the
model galaxy (observable) properties are given in Table 2. The MAH
was quantified through the $\gamma $ parameter, where $\gamma =\frac{%
  M_0-M_0/2}{t(M_0)-t(M_0/2)}.$ It is seen that the MAH, which drives
the SF history, strongly influences B-V, and moderately influences the
gas fraction. The range of B-V colors that the models span is mainly
associated with the statistical dispersion in the MAHs. This
dispersion is also reflected in the TF scatter, and as was shown above
(see Figure 6) observations confirm a correlation between B-V and the
maximum circular velocity for a given luminosity (mass). The $\lambda
$ parameter strongly influences the surface brightness, b/d ratio, and
gas fraction, and slightly influences the color index. For a given $M$
and MAH, $\lambda $ determines the degree of concentration of disks.
According to the SF mechanism used in our models, the stellar disks
formed from more concentrated gaseous disks are typically ``colder''
than those emerged from less concentrated gaseous disks. Therefore,
the stellar Toomre parameter is small for small $\lambda ^{\prime }s$.
That is why the b/d ratio is closely related to $\lambda .$ Mass
strongly correlates with luminosity and the Holmberg radius, and
slightly influences the b/d ratio and the surface brightness (both in
the same direction as suggested by observations). As was pointed out
in \S 3.2, B-V and $\mu _{B_o}$ are the two parameters from which the
other intensive properties depend, i.e. the intensive properties of
disk galaxies may be described in a biparametrical sequence, whose
origin deals with two of the fundamental physical factors of galaxies,
the MAH and $\lambda ,$ respectively. Most of these properties are
almost invariant to the third factor, the mass (luminosity).

\placetable{tbl-2}

The morphological Hubble classification has been a useful guide to
study the observational properties and correlations of galaxies.
Nevertheless, most of the classification discriminators deal with
morphological characteristics that probably are transient phenomena
related to other more fundamental galactic characteristics. The main
classification discriminators for spirals are the pitch angle and the
strength of the spiral arms and bars. The problem of the origin and
maintenance of arms in disk galaxies is however still not well
understood. Within the framework of the density wave theory, feedback
and amplification (overreflection) processes such as the ``waser''
mechanism (Mark 1976) were proposed in order to explain the
self-excitation and persistence of spiral arms without any external
driving, and in the so called modal approach (e.g., Bertin \& Lin
1989a, 1989b and references therein) they are unified in one scheme
which associates the spiral arm structure and the existence of bars
with global modes of oscillation.  Unfortunately, this and other
approaches are not predictive in the sense that they do not provide us
with a direct connection between the structural and dynamical
properties of a given galactic model and the characteristics of its
perturbed state (for example the arm pitch angle). Bertin et al.
(1989a) have applied a global stability analysis to basic states
defined by simple analytical expressions, and found that the generated
survey of models comprises the main morphological types, where three
are the physical parameters which control the morphology: the gas
fraction, the active disk mass with respect to the total mass, and the
``temperature'' of the stellar disk (see also Bertin \& Lin 1996). It
is interesting to note that if the
b/d ratio is understood as a dynamical ``thermometer'' of the system, and $%
\mu _{B_0}$ as an indicator of the disk self-gravity, then according
to the prediction of our models, galaxies fill only a planar region in
the space of the three physical parameters which control the
morphology.

Bertin \& Romeo (1987), Bertin et al. (1989a,b) and other authors (see
for references Combes 1993) coincide in pointing out that gas,
dynamically speaking, is crucial for the excitation and maintenance of
spiral structures. A purely stellar spiral would heat up quickly and
disappear, producing a thickened, ``hot'' structure. {\it Accretion of
  gas, which is natural in the extended collapse scenario, will cool
  the stellar disk and give the conditions for instabilities}. Such a
behavior has also been reported in N-body numerical simulations
(Sellwood \& Carlberg 1984).  According to Bertin \& Lin (1996), the
gas fraction mainly determines the sequence of types a, b, and c, i.e.
the pitch angle of arms. As is seen in Table 1, the gas fractions in
our models correlates with the other secondary indicators of the HS
which go across the a, b, and c sequence, such as $\mu _{B_0},$\ b/d,
and B-V, showing that the origin of the Hubble morphological types is
closely related to the galaxy formation and evolution processes of the
extended collapse scenario.

Several authors have pointed out that for a given luminosity galaxies
with larger maximum rotation velocities (the TF dispersion) are of
earlier types than those with smaller velocities (Roberts 1978; Rubin
et al. 1980; Rubin 1985; Giraud 1986, 1987; Krann-Korteweg, Cameron,
\& Tamman 1988; Giovanelli et al. 1997). The B-band TF relation,
$L_{B_{\odot }}=A_{TF}V_{\max }^{m_B},$ predicted by the models has a
dispersion that, to a first approximation, we express only through
variations in the coefficient $A_{TF}$. As has been commented in \S\ 
3.2 (see also Figure 6), it turns out that the models present
correlations between $A_{TF}$ and some secondary Hubble type
indicators: B-V, $f_g,$ and b/d (see Table 1). Hence, within the
framework of the extended collapse scenario the empirical correlation
found between galactic type and rotation velocity for a given
luminosity finds a natural explanation (see also \S\ 3.2 and Figure
6).

\section{Evolutionary models and intermediate redshift ($z\lesssim 1)$
  observations}

With the advent of observational information for galaxies at different
redshifts, an empirical picture of galaxy formation and evolution
begins to be possible (e.g., Ellis 1998). In order to interpret the
observational data of galaxies at intermediate and high redshifts, the
theoretical models are crucial. The models presented here explain in
principle the main characteristics of disk galaxies corresponding to
the locally dominant galaxy population. Since these are evolutionary
models, it is easy to predict how such a population will evolve.

The SF history (SFH) is a relevant evolutionary feature of galaxies.
Within the framework of our models the SFH is driven by the gas
accretion rate
(related to the MAH) and by the gas surface density (related to $\lambda )$%
. The total mass aggregation rates at different redshifts for systems of $%
5\times 10^{10}M_{\odot },$ $5\times 10^{11}M_{\odot }$, and $5\times
10^{12}M_{\odot }$ with the average MAHs are depicted in Figure
8(a) (SCDM, $\sigma _8=0.57)$. In panel (b) the corresponding disk
SF rates for $\lambda =0.05$ are plotted; for the $5\times
10^{11}M_{\odot }$ system, the SFHs corresponding to the very extended
and early active MAHs, respectively ($\lambda =0.05),$ are also
plotted. As it is seen the MAH clearly influences on the SFH. In
Figure 8(c) one appreciates the influence on SFH of the disk surface
density which depends on $\lambda ;$ the dotted curves correspond to
the SFHs of $5\times 10^{11}M_{\odot }$ systems with the average MAH
but with $\lambda =0.035$ and $\lambda =0.1$ (upper and lower curves,
respectively). Also in this panel are plotted two extreme cases: a
very extended MAH with $\lambda =0.1$ (lower point-dashed curve), and
an early, active MAH with $\lambda =0.035$ (upper point-dashed curve).
These are rare objects. The massive systems corresponding to early,
active MAHs and low $\lambda ,$ might be of particular interest,
because these are luminous, very active SF objects at early times which could
correspond to galaxies seen at very high redshifts (Avila-Reese 1998;
see also Baugh et al. 1997).

\begin{figure}
\plotone{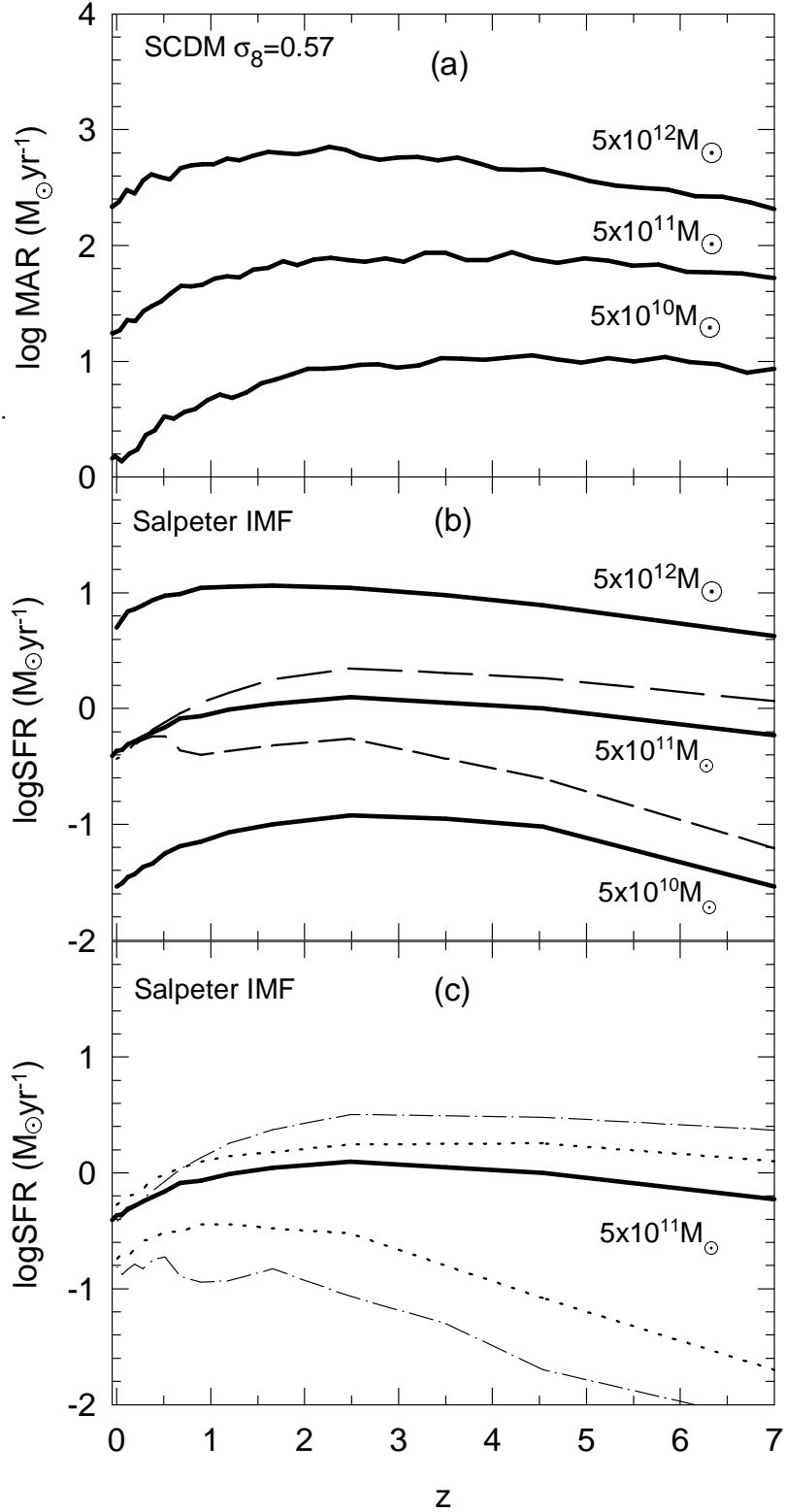}
\caption{In panels (a) and (b) are plotted the total mass
  aggregation rates (MAR), and the SF rates vs. the redshift for
  systems of $5\times 10^{10}M_{\odot },$ $5\times 10^{11}M_{\odot }$,
  and $5\times 10^{12}M_{\odot }$ with the average MAHs and $\lambda
  =0.05,$ respectively.  In panel (b), for systems of $5\times
  10^{11}M_{\odot }$, are also depicted the SFHs corresponding to the
  early active MAH (upper dashed curve), and to
the very extended MAH\ (lower dashed curve). The SFHs for for systems of 
$5\times 10^{11}M_{\odot }$, with the average MAHs, but with $\lambda
=0.035$ (upper dotted curve), and with $\lambda =0.1$ (lower dotted
curve) are ploted in panel (c). Here the upper and lower point-dashed
curves correspond to the extreme cases of an early active MAH with
$\lambda =0.035,$ and a very extended MAH with $\lambda =0.1,$
respectively. \label{fig08}}
\end{figure}

\placefigure{fig08}

According to Figure 8(b), the maximum in the SF activity of the models which
corresponds to a normal disk galaxy population, is attained at $z\sim
1.5-$ $2.5;$ after this, for almost all the cases, the SF rate
moderately decreases until the present epoch by a factor $\sim 2-4$.
Recently, Lilly et al. (1997) have studied the evolution of a large
disk galaxy population using a combination of the observational data
from the CNRS redshift survey and the Hubble Deep Field; they
estimated a moderate decrease in the SF rate, a factor 2.5-3.5 between
$z\approx 0.7$ and $z\approx 0.0$ (for the models, in this case, the
factor is $\sim 2).$ Deep field studies, where no population selection
was made, show that the global (cosmic) SFH per unit
of volume increases from $z\approx 0$ to $z\approx 0.7$ by a factor $\sim 6$  
and by more than a factor of $10$ up to the maximum which is attained at 
$z\approx 1.5-2.0$ (e.g., Madau, Pozzetti, \& Dickinson 1997, and the
references therein). If our models actually describe the evolution of
the normal disk galaxy population which dominates the luminosity
and SF rate today, then the high global SF rate (and luminosity)
observationally estimated at $z\approx 1-2$, can not be produced by
this population.

\begin{figure}
\plotone{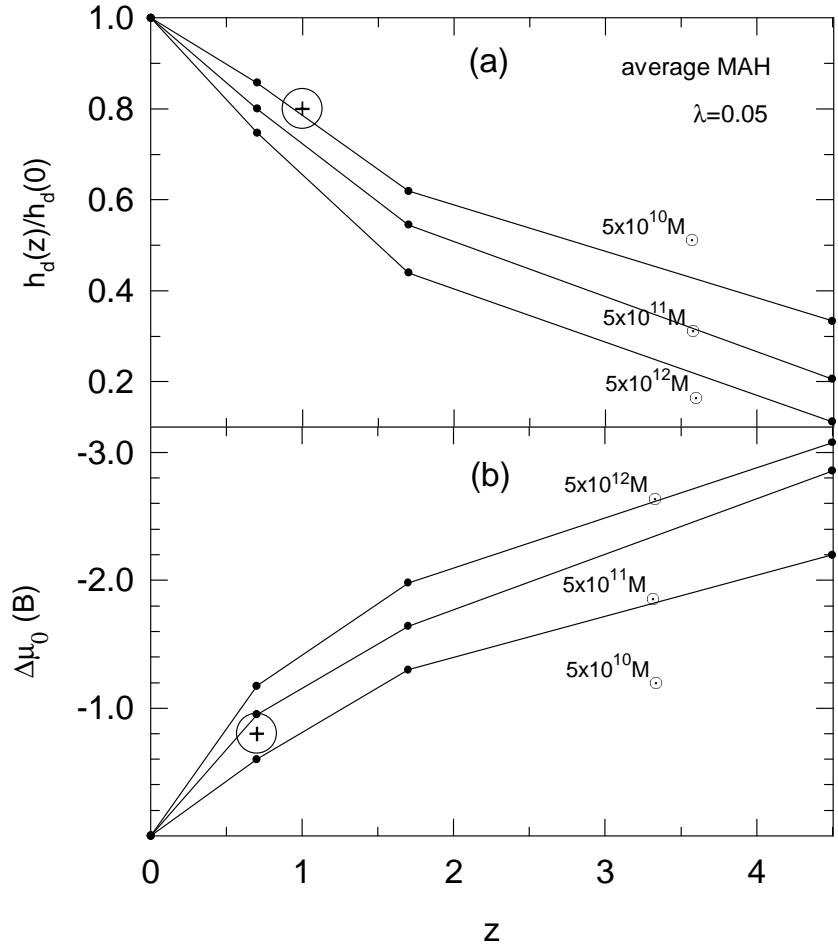}
\caption{Evolution with $z$ of the disk scale radius scaled to 
the radius at $z=0$ (a), and of the difference in mag/arcsec$^2$ of the
central surface B-brightness with respect to its value at $z=0$ (b), for 
models corresponding to systems of $5\times 10^{10}M_{\odot },$ 
$5\times 10^{11}M_{\odot }$, and $5\times 10^{12}M_{\odot }$ with the average
 MAHs and $\lambda =0.05.$ The circles with a cross indicate the corresponding
estimations obtained from deep field observational studies of a bright spiral 
galaxy population (Lilly et al. 1997). \label{fig09}}
\end{figure}

\placefigure{fig09}

In the extended collapse scenario disk galaxies form inside-out by a
continuous process of mass aggregation and are not expected to undergo abrupt
evolutionary changes in their structural and luminosity
characteristics. The evolution of $h_d$, $\mu _{B_o},$ and the
bulge-to-total mass (luminosity) ratio for models corresponding to
$5\times 10^{10}M_{\odot },$ $5\times 10^{11}M_{\odot }$, and $5\times
10^{12}M_{\odot }$ systems with the average MAHs and $\lambda =0.05$
are depicted in Figure 9(a), 9(b), and 10, respectively. Lilly et al.
(1997), for the galaxy population they studied,
estimated a maximum decrease in the scale radius of $25\%$ from $%
z\approx 0$ to $z\approx 1$ (the circle with a cross in panel (a)). At
the same time, from $z\approx 0$ to $z\approx 0.7,$ they
estimated an increase in the central surface brightness (B-band)
corresponding to $\sim 0.9$ mag/arcsec$^2$ (the circle with a cross in
panel (b)). The evolution of the bulge-to-total ratio is less clear.
If we take the average ratios given
in Lilly et al. (1997) at $\left\langle z\right\rangle =0.375$ and $%
\left\langle z\right\rangle =0.625$ we obtain a slope, that after
  normalizing to $z=0,$ corresponds to the segment plotted in Figure
  10. An interesting prediction of our ``secular'' recipe for bulge
  formation is that the less massive systems form their bulges later
  than the more massive systems.

\begin{figure}
\plotone{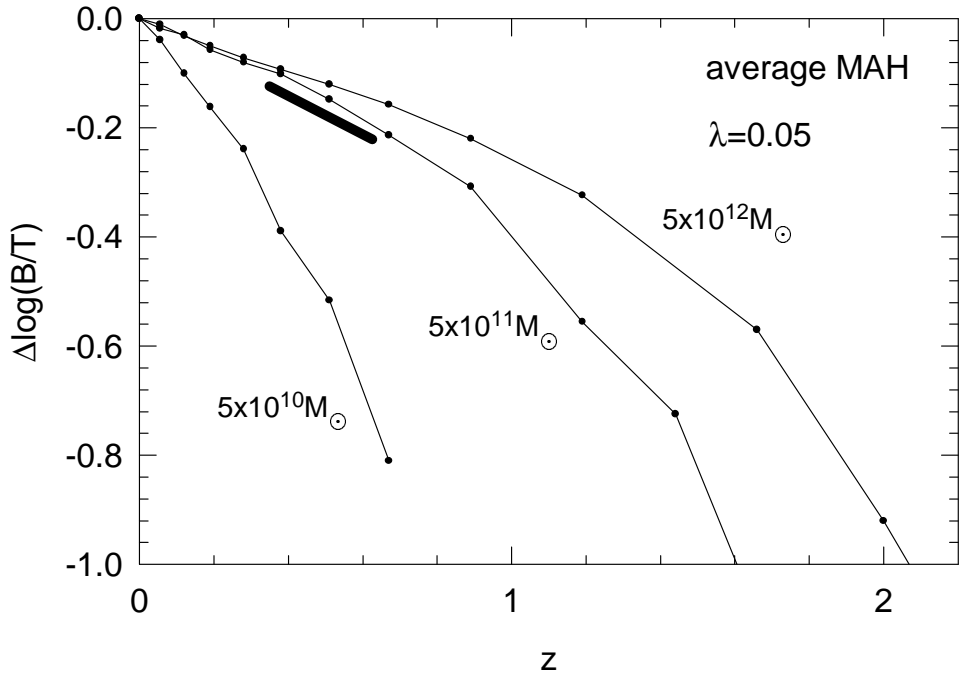}
\caption{Evolution of the bulge-to-total luminosity ratio for 
the same models of figure 9. 
$\Delta \log (b/t)\equiv \log (b/t)(z)-\log(b/t)(0).$ The thick segment
 corresponds to the slope inferred from the observational data presented in 
Lilly et al. 1997 (see text). \label{fig10}}
\end{figure}

\placefigure{fig10}

  The evolution of the B-band TF relation can be related to ({\it i})
  the structural evolution of the galactic system and ({\it ii}) to
  the B-band luminosity evolution. The models (average MAH and
  $\lambda =0.05)$ show that the ``structural'' TF relation,
  $M_s=AV_{\max }^m$ (or equivalently the H- or I-band TF relation),
  has minimal slope changes with $z,$ while the zero-point
  $A$ decreases between $z=0$ and $z=0.7$ by a factor of 2 (0.75 mag)
  and between $z=0$ and $z=1.7$ by a factor of 3 (1.20 mag). The
  luminosity with $z\rightarrow 1.5-2.5$ increases by a factor of
  $2-3$ for most of the models. Regarding the model B-band TF
  relation, its slope also remains approximately the same in the past,
  while the cero point increases with $z$, but not significantly
  ($\sim 0.1$ mag and $\sim 0.45$ mag for the mentioned redshift
  intervals). The B-band TF relation remains almost the same with $z$
  due to the compensation of 2 effects: the structural evolution of
  the system and the luminous evolution of the disk. Vogt et al.
  (1997), from a deep field study (up to $z\approx 0.7)$, have
  concluded that the slope of the B-band TF relation does not change,
  while the zero-point could increase no more than $0.4$ mag.

\section{Conclusions}

We have modeled the formation and evolution of disk galaxies within
the framework of the extended collapse scenario, which is based on the
inflationary CDM models. The gas disks in centrifugal equilibrium were
built-up under the assumption of detailed angular momentum
conservation into spherical virializing dark matter halos whose MAHs
were calculated from the initial cosmological conditions. The disk SF
is produced by global gravitational instabilities and is
self-regulated by an energetic balance of the turbulent gas. The
bulges are formed by secular evolution of the stellar disk based on
 gravitational instabilities. The main predictions of the models
are:

1). The disks present exponential surface brightness profiles and
negative radial B-V gradients. The scale lengths and central surface
brightnesses are in agreement with the observations, including the LSB
galaxies.

2). The rotation curves are nearly flat up to the Holmberg radius.
Contrary to observational estimations, the rotation curve
decompositions show dominion of dark matter down to the galaxy central
regions. A constant density core in the dark halo solves this problem.

3). The intensive properties and their correlations (particularly
those which go across the HS) of the models corresponding to the local
($z\approx 0)$ population of disk galaxies, including the LSB
galaxies, are determined by the combination of three fundamental physical
factors and their statistical distributions, related to the initial
cosmological conditions. These three factors are the mass, the MAH, and
the primordial angular momentum expressed through the spin parameter
$\lambda .$

4). The intensive properties of the models can be described in a
biparametrical sequence, where the parameters may be the color index B-V
and the central surface brightness $\mu_{B_o}$. Each one of these parameters is
determined mainly by the MAH and $\lambda $, respectively. The third
fundamental physical factor, the mass, exerts no practical influence
the intensive properties. We have shown that the empirical
luminosity (mass)-color relation (or equivalently the color TF
relation) can be explained by the effects of the metallicity and the 
extinction.observed dependence of extinction on
luminosity (mass). These effects also contribute to decrease the slope of the
B-band TF relation.

5). The SF rates of models with the average MAHs and $\lambda =0.05$
grow by factors of 2.5-4.0 up to $z\sim 1.5-2.5$ with respect to the
SF rates at $z=0 $. After this maximum, the SF rates slowly decrease
with $z.$ The SFHs of systems with early, active MAH and/or low
$\lambda ^{\prime }s$ show high SF rates at high redfshifts ($z>3),$
while the systems with extended MAHs and/or high $\lambda ^{\prime }s$
present small SF rates which slowly increase until the present epoch.

6). The structural properties of the models do not change abruptly.
Between $z=0$ and $z\approx 1$ the disk scale radii in average
decrease a factor $\sim 1.3$ and the central surface brightnesses
increase $\sim 1$ mag/arcsec$^2.$ The bulge-to-total luminosity ratio
also decreases with $z$ and decreases more severely for the low mass systems.
The slopes of the ``structural'' and B-band TF relations do not change
with $z.$ In the case of the ``structural'' TF relation, the
zero-point decreases (0.75 mag for $z=0.7$ with respect to $z=0$),
while for the B-band TF relation the zero-point slightly increases
(0.1 mag at $z=0.7$ with respect to $z=0).$

The exploratory models presented in this work show that the main
observational characteristics and correlations of disk galaxies can be
well understood in the context of the extended collapse scenario,
suggesting a direct connection between the conditions prevailing in
the early universe and the properties of galaxies today. A serious
shortcoming of galaxies emerging from Gaussian CDM cosmological models
is the gravitational dominion of DM over baryon matter. The
remedy to this problem is the introduction of a core in the DM halo.
Fortunately, the intensive galaxy properties and their correlations
are not significantly sensitive to the existence or non-existence of such a
core, in such a way that all the results presented here are also true
for galaxies with a core in their DM halos. The main limitations of
our approach are connected to the facts that {\it (i)} the influence
of the environment on galaxy formation and evolution was not taken
into account, {\it (ii)} detailed angular momentum conservation for the
baryon gas collapse was assumed, and {\it (iii)} the mass aggregation
was treated only as gas accretion neglecting the possibility of
mergers of stellar systems. In future we shall address ways of
overcoming these limitations with the aim to improve the model
predictions and to explain galactic properties and distributions
related to environment.

\acknowledgments

V.A.-R. acknowledges the support of a fellowship from the project UNAM-DGAPA
IN-105894 and support from the program ``Becas Cuauht\'{e}moc'' of CONACyT. 
C.F and V.A.-R.  would like to acknowledge to Antonio Garc\'{e}s for his
computing assistence.

\begin{center}
  References
\end{center}

\begin{description}
\item Avila-Reese, V. 1998, PhD. Thesis, U.N.A.M (in spanish).

\item  Avila-Reese, V., Firmani, C., \& Hern\'{a}ndez, X. 1997, \apj, accepted (astro-ph/9710201) (AFH)

\item Baugh, C.M., Cole, S., \& Frenk, C.S. 1996, \mnras, 283, 1361

\item Baugh, C.M., Cole, S., \& Frenk, C.S., \& Lacey, C. 1997,
  submitted to \apj, (astro-ph/9703111)

\item Begeman, K. 1987, Ph.D. Thesis, University of Groningen

\item Bertin, G., \& Lin, C.C. 1996, ``Spiral Structure in Galaxies. A
  Density Wave Theory'' (The MIT Press)

\item Bertin, G., \& Romeo, A.B. 1988, \aap, 195, 105

\item Bertin, G., Lin, C.C., Lowe, S.A., \& Thurstans, R.P. 1989a,
  \apj, 338, 78

\item \_\_\_\_\_. 1989b, \apj, 338, 104

\item Boselli, A., \& Gavazzi, G. 1994, \aap, 283, 12

\item Bressan, A., Chiosi, C., \& Fagotto, F. 1994, \apjs, 94, 63

\item Burkert, A. 1995, \apj, 447, L25

\item Bruzual, G.A., \& Charlot, S. 1993, \apj, 405, 538

\item Cardelli, J.A., Clayton, G.C., \& Mathis, J.S. 1989, \apj, 345,
  245

\item Carignan, C., \& Freeman, K.C. 1985, \apj, 160, 811

\item Carlberg, R.G. 1985, in ``The Milky Way Galaxy'', proceedings of
  the 106th Symposium (Dordreicht, D.Reidel Publsihing Co.), 615

\item Catelan, P., \& Theuns, T. 1996, \mnras, 282, 436

\item Charlot, S., Worthey, G., \& Bressan A. 1996, \apj, 457, 625

\item Cole, S., Aragon-Salamanca, A., Frenk, C.S., Navarro, J., \&
  Zepf, S.  1994, \mnras, 271, 781

\item Combes, F. 1993, in ``The Formation and Evolution of Galaxies'',
  eds.  Mu\~{n}oz-Tu\~{n}\'{o}n \& F. S\'{a}nchez (Cambridge Univ.
  Press), p. 317

\item Courteau, S., de Jong, R.S., \& Broeils, A.H. 1997, \apj, 457,
  L73

\item de Jong, R.S. 1995, Ph.D. Thesis, University of Groningen

\item \_\_\_\_\_. 1996a, \aap, 313, 45

\item \_\_\_\_\_. 1996b, A\&ASS, 118, 557

\item de Vaucouleurs, G., et al. 1991, Third Reference Catalogue of
  Bright Galaxies (Springer, Berlin Heidelberg New York)

\item Dekel, A., \& Silk, J. 1986, \apj, 303, 39

\item Disney, M.J., Davies, J., \& Philipss, S. 1989, \mnras, 205,
  1253

\item Ellis, R.S 1998, in ``Cosmological Parameters \&\ Evolution of
  the Universe'', ed. K.Sato, IAU Symposium 183, in press

\item Gallagher, J.S., Hunter, D.A., \& Tutukov, A.V. 1984, \apj, 284,
  544

\item Firmani, C., \& Tutukov, A.V. 1992, \aap, 264, 37

\item \_\_\_\_\_. 1994, \aap, 288, 713

\item Firmani, C., Hern\'{a}ndez, X., \& Gallagher 1996, \aap, 308,
  403

\item  Flores, R.A., Primack, J.R., Blumenthal, G.R., \&\ Faber, S.M. 1993, %
  \apj, 412, 443

\item Fritze-v.Alvensleben, U., \& Gerhard, O.E. 1994, \aap, 285, 751

\item Gavazzi, G. 1993, \apj, 419, 469

\item Giovanelli, R., Haynes, M.P., Herter, T.H., Vogt, N.P., da
  Costa, L.N., Freudling, W., Salzer, J.J., \& Wegner, G. 1997, \aj,
  113, 53

\item Giovanelli, R., Haynes, M., Salzer, J.J., Wegner, G., Da Costa,
  L.N., Freudling, W. 1995, \aj, 110, 1059

\item Giraud, E. 1986, \apj, 309, 512

\item \_\_\_\_\_. 1987, \aap, 174, 23

\item Gunn, J.E. 1981, in ``Astrophysical Cosmology'', M.S. Longair,
  Coyne G.V., \& H.A. Br\"{u}ck, eds. (Pontificia Academia
  Scientarium:Citta del Vaticano), p.191

\item \_\_\_\_\_. 1987, in ``The Galaxy'', G. Gilmore \& B. Carswell,
  eds.  (Reidel Publishing Company), p.413

\item Heavens, A., \& Peacock, J. 1988, \mnras, 232, 339

\item Hoffman, Y. 1986, \apj, 301, 65

\item \_\_\_\_\_. 1988, \apj, 329,8

\item Kauffmann, G. 1995, \mnras, 274, 161

\item \_\_\_\_\_. 1996, \mnras, 281, 475

\item Kauffmann, G., White, S.D.M., \& Guiderdoni, B. 1993, \mnras,
  264, 201

\item Kennicutt, R.C. 1983, \apj, 272, 5

\item \_\_\_\_\_. 1989, \apj, 344, 6854

\item Kennicutt, R.C., Tamblyn, P., \& Congdon, C.W. 1994, \apj, 435,
  22

\item Kodama, T., \& Arimoto, N. 1997, \aap, 320, 41

\item Kraan-Korteweg, R.C., Cameron, L.M., \& Tammann, G.A. 1988,
  \apj, 331, 620

\item Kudrya, Yu.N., Karachentseva, V.E., Karachentsev, I.D, \&
  Parnovsky, S.L. 1997, Pis'ma v Astronomicheskii Jurnal, 23, 730 (in
  Russian)

\item  Lacey, C.G., Guiderdoni, B., Rocca-Volmerange, \& B.Silk, J. 1993, %
  \apj, 402, 15

\item Larson, R.B. 1992, in ``Star Formation in Stellar Systems'',
  G.Tenorio-Tagle, M.Prieto, \& F.S\'{a}nchez, eds. (Cambridge
  University Press), p.125

\item Larson, R.B., \& Tinsley, B. M. 1978, \apj, 219, 46

\item Larson, R.B., \& Tinsley, B. M., \& Caldwell, C.N. 1980, \apj,
  237, 692

\item Lilly, S.J. et al. 1997, preprint (astro-ph/9712061)

\item Madau, P., Pozzetti, L., \& Dickinson, M. 1997, preprint
  (astro-ph/9708220)

\item Mark, J.W.-K. 1976, \apj, 205, 363

\item McGaugh, S.S., de Blok, W.J.G. 1997, \apj, 481, 689

\item Norman, C.A., Sellwood, J.A., \& Hassan, H. 1996, \apj, 462, 114

\item Nulsen, P.E.J., \& Fabian, A.C. 1995, \mnras, 277, 561

\item Peletier, R., \& Balcells, M. 1996, \aj, 111, 2238

\item Persic, P., Salucci, M., \& Stel, F. 1996, \mnras, 281, 27

\item Renzini, A. 1994, in ``Galaxy Formation'', J.Silk \& N.
  Vittorio, eds. (Elsevier Science Publishers B.V., The Netehrlands),
  p.303

\item Rees, M.J., \& Ostriker, J.P. 1977, \mnras, 179, 541

\item Roberts, M.S. 1978, \aj, 83, 1026

\item Roberts, M.S., \& Haynes, M.P. 1994, \araa, 32, 115

\item Sandage, A. 1961, The Hubble Atlas of Galaxies (Carnegie Inst.
  of Washington, Washington DC)

\item Rubin, V.C., Burstein, D., Ford, W.K., \& Thonnard, N. 1980,
  \apj, 238, 471

\item \_\_\_\_\_. 1985, \apj, 289, 81

\item Ryden, B.S., \& Gunn, J.E. 1987, \apj, 318, 15

\item Sellwood, J.A., \& Carlberg R.G. 1984, \apj, 282, 61

\item Silk, J. 1977, \apj, 211, 638

\item Simien, F., \& de Vaucouleurs, G. 1986, \apj, 302, 564

\item Skillman, E.D. 1987, in ``Star Formation in Galaxies'', C.J.
  Lonsdale, ed. (NASA Conference Publication N0. 2466, Washington),
  p.263

\item Strauss, M.A., \&\ Willick, J.A. 1995, \physrep, 261, 271

\item Tinsley, B. 1980, Fundamental Cosmic Phys., 5, 287

\item Toomre, A. 1964, \apj, 139, 1217

\item T\'{o}th, G., \& Ostriker, J.P. 1992, \apj, 389, 5

\item Tully, R.B. 1988, Nearby Galaxies Catalog (Cambridge, U.P.)

\item Tully, B., Mould, J., \& Aaranson, M. 1982, \apj, 257, 527

\item Vogt, N. et al. 1997, \apj, 479, L121

\item Vishnavatan, N. 1981, \aap, 100, L20

\item Wang, B., \& Heckman, T.M. 1996, \apj, 457, 645

\item White, S.D.M. 1994, preprint MPA 831

\item White, S.D.M, \& Frenk, C.S. 1991, \apj, 379, 52

\item White, S.D.M., \& Rees, M.J. 1978, \mnras, 183, 341

\item White, S.D.M., Efastathiou, G,. \& Frenk, C.S. 1993, \mnras,
  262, 1023

\item Wyse, R. 1982, \mnras, 199, 1P

\item Wyse, R.F.G., Gilmore, G., \& Franx, M. 1997, \araa, 35, 637
\end{description}

\begin{deluxetable}{lrrrrrrr}
  \tablenum{1}
\tablewidth{0pt}
\tablecaption{Correlation matrix of the global properties \label{tbl-1}}
\tablehead{ \colhead{ } & \colhead{B$$-$$V} & \colhead{$f_g$} &
  \colhead{b/d} & \colhead{$M_B$} & \colhead{$V_m$} & \colhead{log
    $h_d$} & \colhead{$A_{TF}$}} \startdata $\mu_{B_0}$ & $-$0.07&
0.84& $-$0.93& 0.55& $-$0.65& 0.29& 0.33 \nl B$$-$$V & \nodata &
$-$0.56& 0.18& 0.22& $-$0.05& $-$0.29& $-$0.82 \nl $f_g$ & \nodata &
\nodata & $-$0.87& 0.27& $-$0.45& $-$0.01& 0.68 \nl b/d & \nodata &
\nodata & \nodata & $-$0.38& 0.54& 0.10& $-$0.51 \nl $M_B$ & \nodata &
\nodata & \nodata & \nodata & $-$0.94& $-$0.95& $-$0.20 \nl $V_m$ &
\nodata & \nodata & \nodata & \nodata & \nodata & 0.84& $-$0.04 \nl
log $h_d$ & \nodata & \nodata & \nodata & \nodata & \nodata & \nodata
& 0.34 \nl

\enddata
\end{deluxetable}


\begin{deluxetable}{lrrrrrrrr}
\tablenum{2}
\tablewidth{0pt}
\tablecaption{Correlation matrix of the global properties and the fundamental parameters \label{tbl-2}}
\tablehead{
\colhead{    }           & \colhead{$\mu_{B_0}$} &
\colhead{B$-$V}          & \colhead{$f_g$}       &
\colhead{b/d}            & \colhead{$M_B$}       &
\colhead{$V_m$}          & \colhead{log $h_d$}   &
\colhead{$A_{TF}$}}
\startdata
$\log M_0$\tablenotemark{(a)} & $-$0.44 & $-$0.24 & $-$0.17 &    0.27 & 
$-$0.9 & 0.92 & 0.98 & 0.25 \nl
$\gamma$\tablenotemark{(b)}  &    0.01 & $-$0.95 &    0.48 & $-$0.12 & 
$-$0.24& 0.05 & 0.29 & 0.82 \nl
$\lambda $\tablenotemark{(c)} &    0.86 & $-$0.07 &    0.81 & $-$0.88 &   
0.13& $-$0.23 & 0.51 & 0.34 \nl

\tablenotetext{(a)}{present-day mass}
\tablenotetext{(b)}{$\gamma =\frac{M_0-M_{0/2}}{t(M_0)-t(M_{0/2})}$}
\tablenotetext{(c)}{spin parameter}
\enddata
\end{deluxetable}

\clearpage

\end{document}